\documentclass[preprint,hidelinks,olorlinks=true,linkcolor=blue,citecolor=blue,urlcolor=blue]{aastex63}

\received{received date}
\revised{revised date}
\accepted{accepted date}
\submitjournal{ApJ}

\shorttitle{2D Clouds on Tidally Locked Planets}

\shortauthors{Song, Yang, Luo, Li, and Fu}

\graphicspath{{./}{figures/}}
\usepackage{float}
\usepackage{multirow}

\begin{document}
\title{Idealized 2D Cloud-Resolving Simulations for Tidally Locked Habitable Planets}

\correspondingauthor{Jun Yang}
\email{junyang@pku.edu.cn}

\author{Qiyu Song}
\affiliation{Department of Atmospheric and Oceanic Sciences, School of Physics, Peking University, Beijing 100871, China.}

\author{Jun Yang}
\affiliation{Department of Atmospheric and Oceanic Sciences, School of Physics, Peking University, Beijing 100871, China.}

\author{Hang Luo}
\affiliation{Department of Atmospheric and Oceanic Sciences, School of Physics, Peking University, Beijing 100871, China.}

\author{Cheng Li}
\affiliation{Department of Climate and Space Sciences and Engineering, University of Michigan, Michigan 48109, USA}

\author{Shizuo Fu}
\affiliation{Key Laboratory for Humid Subtropical Eco-Geographical Processes of the Ministry of Education, Fujian Normal University, Fuzhou 350007, China}
\affiliation{School of Geographical Sciences, Fujian Normal University, Fuzhou 350007, China}



\begin{abstract}

Cloud is critical for planetary climate and habitability, but it is also one of the most challenging parts of studying planets in and beyond the solar system. Here we use a cloud-resolving model (CRM) with high resolution (2 km) in a two-dimensional (2D) configuration to simulate the clouds and circulation on tidally locked aqua-planets. We find that the substellar area is covered by deep convective clouds, the nightside is dominated by low-level clouds, and these two are linked by a global-scale Walker circulation. We further find that a uniform surface warming causes the substellar cloud width to decrease, but a reduction in day-night surface temperature contrast or an increase in longwave radiative cooling rate causes the substellar cloud width to increase. These relationships can be roughly interpreted based on simple thermodynamic theories. \textcolor{black}{Comparing the results between CRM and global 3D general circulation model (GCM), we find that they show qualitatively consistent results, including the Walker circulation, the substellar clouds, and the responses of the substellar ascending area and strength to changes in surface temperature or in its zonal contrast. But, large quantitative differences exist, such as the magnitude of cloud water path, the cloud width, and their responses to external forcings. These results increase our confidence in using GCMs for modeling exoplanetary climates, although large quantitative uncertainties should always exist.} Future work is required to use 3D CRM(s) with realistic radiative transfer and with the Coriolis force to examine the clouds and climate of tidally locked planets.

\end{abstract}


\section{Introduction} \label{sec:intro}
Tidally locked terrestrial planets around M and K dwarfs are the most probable candidates for finding potentially habitable worlds beyond Earth. The lifetime of the host stars is longer than other main-sequence stars, and the distances between potentially habitable planets and the stars are relatively small, thus the transits of these planets are more frequent and significant, which means more opportunity to observe them. In this study, we focus on the climate simulations of tidally locked habitable planets, in particular on convection and clouds.

Using atmospheric general circulation models (GCMs) or coupled atmosphere--ocean GCMs, previous climate simulations showed that the dayside of 1:1 tidally locked (or called synchronously rotating) habitable planets would be covered by deep convective clouds, and the nightside would be covered by low-level stratus clouds. The two hemispheres are connected with each other through a global-scale atmospheric overturning circulation, or called global-scale Walker circulation, with strong upwelling and low-level convergence over the substellar region and weak downwelling and low-level divergence in the rest of the planetary atmosphere \citep[e.g.,][]{joshi1997simulations,merlis2010atmospheric,Wordsworth_2010,edson2011atmospheric,yang2013stabilizing,menou2013water,kopparapu2016inner,kopparapu2017habitable,wolf2017assessing,salameh2017climate,boutle2017exploring,Haqq-Misra2018demarcating,Yang_2019_ocean,del2019habitable}. The atmospheric circulation is largely driven by the continuous stellar heating on the permanent dayside and the longwave cooling \textcolor{black}{on both dayside and nightside, especially for slow rotators}. The dayside convective clouds strongly reflect the stellar radiation, increase the planetary albedo, and thereby cool the surface. These clouds make the inner edge of the habitable zone be closer to the host stars, expanding the width of the habitable zone. \textcolor{black}{The rotation rate of the planet can also strongly affect the climate \citep[e.g.,][]{kopparapu2017habitable, Haqq-Misra2018demarcating}. For fast rotators, clouds are stretched towards the nightside and form a band along the equatorial area, and thus their planetary albedo is relatively lower than those of slowly rotating planets. Faster rotation rates can also lead to smaller day-night temperature contrast due to stronger equatorial superrotation and more effective energy transport from the dayside to the nightside.}

However, GCMs have some essential shortcomings, mainly rising from their low resolution, which is on the order of hundreds of kilometers in the horizontal direction. As the low resolution is insufficient to simulate the convection and cloud processes explicitly, GCMs usually employ parameterization schemes for convection and clouds, which are based on observations and cloud-resolving simulations of Earth. Whether these parameterization schemes could be applied to exoplanets' conditions is unknown, and different models employ different parameterization schemes. Recent model intercomparisons for tidally locked planets \citep{yang2019simulations,fauchez2020trappist,sergeev2021trappist} showed that there are large differences among the models in simulating the climate. Under given stellar flux, given greenhouse gas concentration and the same boundary conditions (such as a uniform slab ocean at the surface), the simulated surface temperature difference in global mean could be as large as 20 K to 30 K. The underlying reasons are mainly from cloud parameterization, radiation calculation accuracy for water vapor\textcolor{black}{, and differences in} relative humidity (which is associated with large-scale atmospheric circulation)\textcolor{black}{; the} largest difference comes from clouds \citep{Yang_2016,yang2019simulations}. Improving our understanding of clouds and their effects on climate on tidally locked planets is in urgent need.

Two ways could be used to improve the studies of clouds on tidally locked habitable planets. One is to observationally identify clouds on terrestrial exoplanets, but it is far beyond the ability of present and near-future space and ground telescopes. The other one is to resolve the cloud-formation processes in exoplanets' atmospheres, which is considered in this study. Until now, there are three cloud-resolving studies for tidally locked planets around M dwarfs, \cite{zhangetal2017surface}, \cite{sergeev2020atmospheric}, and \cite{Lef_vre_2021}. Due to computation resource limits, only a limited-area domain is cloud-resolved, $\approx$1000 by 1000 km in \cite{zhangetal2017surface}, 6000 by 6000 km in \cite{sergeev2020atmospheric}, and 250 by 250 km in \cite{Lef_vre_2021}. \cite{zhangetal2017surface} found that there are strong spatial variability in clouds, shortwave radiation, and surface temperature even in the small domain. This is likely due to the small scales of deep convection. \cite{sergeev2020atmospheric} examined the dependence of the simulated climate on the choice of convection parameterization scheme and a cloud-permitting modeling of the substellar convection. They found that the simulated clouds, precipitation, atmospheric circulation, and day-to-night energy transport are sensitive to how convection and clouds are treated. They also showed that the planetary albedo of the substellar region in the cloud-permitting simulation is somewhat higher than that in the simulation with parameterized convection and clouds. \cite{Lef_vre_2021} confirmed that the substellar region of tidally locked habitable planets should be covered by convective clouds. They also found that there are large differences in the simulated cloud albedo between the CRM and GCMs, due to less formation of low-altitude stratocumulus clouds in the CRM they employed.  \textcolor{black}{This study is a similar or further step of the previous three studies using a different cloud-resolving model. Specially, we will examine the separated effects of changing the surface temperature uniformly and changing the day-night surface temperature contrast, and we will compare the CRM results with GCM results. Moreover, simple thermodynamical theories including moist static energy (MSE) and weak temperature gradient (WTG) are employed in order to more clearly understand the results.}



In the cloud-resolving experiments for Earth, the horizontal resolution is always several kilometers or even less, and the time step is always in the order of seconds \citep[e.g.,][]{miller1974three,klemp1978simulation,tao1986study,tomita2005global,muller2012detailed,wofsy2012cloud}, both of which are at least two orders smaller than those in GCMs, so lots of computation resources as well as storage disks are required. Due to the very high computation resource requirement, cloud-resolving simulations always employ a 2D configuration or a regional 3D configuration \citep[e.g.,][]{muller2012detailed,wofsy2012cloud}. Even for Earth, only few studies have employed global or near-global cloud-resolving simulations \citep[e.g.,][]{tomita2005global,bretherton2015convective,Khairoutdinov2018,Narenpitak2019,stevens2019dyamond}. As a further step using the method of cloud resolving, we simulate the clouds on tidally locked habitable planets in a 2D configuration (longitude--height) but in a large zonal (west--east) domain, 40,000 km, encircling the entire equator of the planet. The domain is large enough to consistently investigate the interaction between convection/clouds and large-scale circulation, but the meridional (south--north) direction is absence in this study.


Previous studies showed that for the Walker Circulation over the Pacific Ocean on Earth, a 2D model configuration along the equator is very useful in studying the effects of sea surface temperature, surface temperature gradient, radiative transfer, cloud microphysics, and other factors on the sensitivities of the Walker Circulation and clouds \citep{grabowski2000cloud,bretherton2002simple,bretherton2006interpretation,wofsy2012cloud,Liu2008explicitly,byrne2016energetic,chen2019role}. As the atmospheric circulation on tidally locked terrestrial planets has certain similarities with the Walker circulation on Earth \citep[e.g.,][]{joshi1997simulations,edson2011atmospheric,merlis2010atmospheric,yang2013stabilizing}, we employ a similar 2D model configuration. On Earth, the main upwelling occurs over the warm pool of the western Pacific, and the main downwelling occurs over the cold tongue of the eastern Pacific. On tidally locked planets, the upwelling is mainly over the substellar area and the downwelling over the rest of the planet. But there are two main differences between our configuration and the configurations widely used for the Walker Circulation simulations on Earth: one is that the spatial scale is larger, 40,000 km for a whole circle along the equator versus $\approx$20,000 km for the width of the Pacific ocean, and the other one is that the zonal (west--east) temperature difference is tens of degrees versus several degrees in K or $^\circ$C. Therefore, the results shown here are more applicable for tidally locked planets than for the regional Walker Circulation on Earth.

The structure of this paper is as follows. In Section \ref{sec:methods}, we introduce the CRM used and address the experimental designs. Corresponding GCM experimental designs are also shown. In Section \ref{sec:results}, we first show the general characters of the cloud pattern and the large-scale circulation. Then, we study the effects of different factors on the clouds and circulation, including making uniform surface warming or cooling, varying the day-night surface temperature contrast, and changing the strength of atmospheric radiative cooling, and then explore the underlying mechanisms (subsections \ref{subsec:mech_body} and \ref{subsec:mech_anvil}). The \textcolor{black}{results of GCM experiments} are shown in subsection \ref{subsec:GCM_results}. Finally, we \textcolor{black}{discuss several related topics in Section \ref{sec:discussion}, and} draw the summary in Section \ref{sec:sum}.



\section{Model Descriptions and Experimental Designs} \label{sec:methods}

\subsection{2D CRM Simulations} \label{subsec:CRM_method}
The model used here is the Simulating Nonhydrostatic Atmospheres of Planets (SNAP, \cite{Li2019Simulating}, link: \url{https://snap.chengcli.io/}), a nonhydrostatic CRM, developed for performing high-resolution studies including moist convection and cloud processes in planetary atmospheres of various types of planets. The model employs the finite volume method to solve the hydrodynamic equations including density, velocities, vapor, cloud, precipitation, and total energy, because this method maintains conservation quantities and deals well with variables with large gradients. The equations of the model are originally written in a 3D space, but one horizontal dimension ($y$ direction) is left out in this work because we perform simulations in a 2D space.

The continuity equations of different components in the atmosphere are written as:
\begin{eqnarray}
\frac{\partial \rho_{i}}{\partial t}+\frac{\partial\left(\rho_{i} u\right)}{\partial x}+\frac{\partial\left(\rho_{i} w\right)}{\partial z} & = & 0, \quad i=d,v,c \\
\frac{\partial \rho_{p}}{\partial t}+\frac{\partial\left(\rho_{p} u\right)}{\partial x}+\frac{\partial\left(\rho_{p} w\right)}{\partial z} & = & -\frac{\partial\left(\rho_{p} \overline{w}_{p}\right)}{\partial z},
\end{eqnarray}
where $\rho_d, \rho_v, \rho_c,$ and $\rho_p$ represent the density of dry air, water vapor, cloud, and precipitation, respectively. $(u, w)$ are the 2D velocity components in $(x, z)$ directions. $\overline{w}_p$ is the terminal velocity of precipitation with respect to the mean velocity of the background atmosphere and is set to $-10~\mathrm{m\, s^{-1}}$ in all cases, where minus value means moving downwards.

The difference in vertical velocity between precipitation and the background atmosphere also contributes to vertical momentum and energy fluxes. The momentum equations are written as:
\begin{eqnarray}
\frac{\partial(\rho u)}{\partial t}+\frac{\partial(u \rho u)}{\partial x}+\frac{\partial(w \rho u)}{\partial z} & = & -\frac{\partial p}{\partial x} -\frac{\partial\left(\overline{w}_{p} \rho_{p} u\right)}{\partial z} \\
\frac{\partial(\rho w)}{\partial t}+\frac{\partial(u \rho w)}{\partial x}+\frac{\partial(w \rho w)}{\partial z} & = & -\frac{\partial p}{\partial z} -\rho g-\frac{\partial\left(\overline{w}_{p} \rho_{p} w\right)}{\partial z},\label{eq:z-mom}
\end{eqnarray}
where $\rho=\rho_d+\rho_v+\rho_c+\rho_p$ is the total density, and $p=\rho_d R_d T + \rho_v R_v T$ is the total pressure of water vapor and dry air, which are both treated as idealized gases.

The total energy of an air parcel with heterogeneous components consists of internal energy, kinetic energy, and chemical potential energy. The chemical potential energy marks the change in latent heat in phase transitions and is set to zero for gases as a benchmark. The total energy per volume $(\rho e)$ is written as:
\begin{equation}
    \rho e=\rho_{d} c_{v, d} T+\rho_{v} c_{v, v} T+(\rho_{c}+\rho_{p}) c_{l} T + \frac{1}{2} \rho\left(u^{2}+w^{2}\right)+ \mu_{v} \rho_{v}+\mu_{l} (\rho_{c}+\rho_{p}),
\end{equation}
where $c_{v,d}, c_{v,v}$, and $c_{l}$ represent the heat capacity at constant volume of dry air, water vapor, and liquid water (including cloud and precipitation water), respectively. $\mu_{v}$ represents the chemical potential of vapor water, and $\mu_{l}$ represents that of liquid water, and both of them change with air temperature $(T)$. Gravity potential energy is not included in the above equation; instead, the gravity force is treated as an external force to the parcel in Equation~(\ref{eq:z-mom}).

From the continuity equations and momentum equations we can derive the energy equation:
\begin{equation}
\frac{\partial(\rho e)}{\partial t}+\frac{\partial(u\rho e)}{\partial x}+\frac{\partial(w\rho e)}{\partial z} = -\rho g w-\rho_{p} g \overline{w}_{p}-\frac{\partial(up)}{\partial x}-\frac{\partial(wp)}{\partial z}-\frac{\partial\left(\overline{w}_{p} \rho_{p} e_{p}\right)}{\partial z} + \rho \overline{c} \dot{Q},
\label{eq:energy}
\end{equation}
where $e_p=c_{l} T+\frac{1}{2}\left(u^{2}+w^{2}\right)+\mu_{l}$ is the specific total energy of precipitation, and \textcolor{black}{$\overline{c}=(\rho_{d} c_{v, d}+\rho_{v} c_{v, v}+(\rho_{c}+\rho_{p}) c_{l})/\rho$} is the average heat capacity of every component.

\textcolor{black}{The model diagnoses air temperature from the total energy, and the temperature evolves freely under the effects of horizontal and vertical energy transports and the diabatic forcing term ($\dot{Q}$)  with the unit of $\mathrm{K\,s^{-1}}$. The $\dot{Q}$ term includes the radiation heating or cooling of the atmosphere, the stratosphere temperature adjusting towards a reference temperature, and the surface sensible and latent heat fluxes. Among them, only the radiation is prescribed, and others are all dynamically calculated during the model run. Moreover, convective latent heating is not included in $\dot{Q}$, since we have already included the chemical potential in the total energy, which can represent the energy difference between gas and liquid phases.}


The \textcolor{black}{bulk physics} processes in the hydrological cycle are described by the Kessler scheme \citep{kessler1969distribution} written as:
\begin{eqnarray}
\frac{\partial q_{v}}{\partial t} & = & -k_{1}\left(q_{v}-q_{v}^{*}\right)^{+}+k_{4} q_{p}\left(q_{v}^{*}-q_{v}\right)^{+}\label{eq:micro_1} \\
\frac{\partial q_{c}}{\partial t} & = & k_{1}\left(q_{v}-q_{v}^{*}\right)^{+}-k_{3} q_{c} q_{p}-k_{2} q_{c} \\
\frac{\partial q_{p}}{\partial t} & = & -k_{4} q_{p}\left(q_{v}^{*}-q_{v}\right)^{+}+k_{3} q_{c} q_{p}+k_{2} q_{c}.\label{eq:micro_3}
\end{eqnarray}
where $q_v$, $q_c$, and $q_p$ represent mass mixing ratios of water vapor, cloud water, and precipitation water, respectively. Note that the ice phase of clouds is not included in this model. The $k_1$ term is the condensation of over-saturated water vapor and is a rapid process that can be treated as an instant process. The rate of the condensation is proportional to the over-saturated mixing ratio $(q_v-q_{v}^{*})^+$, where $q_{v}^{*}$ is the saturation water vapor mixing ratio, and the expression of $(\cdot)^+$ means taking the maximum of the in-bracket term and zero, i.e., $\max(\cdot,\, 0)$. The $k_2$ term is the auto-conversion from cloud water to precipitation water. The $k_3$ term is the accretion of cloud and rain droplets, which is taken as zero in our simulations. The $k_4$ term is the re-evaporation term of precipitation back to vapor when a raindrop enters a sub-saturate air parcel. We performed two series of sensitivity tests of the \textcolor{black}{bulk physics} parameters $k_2$ and $k_4$ in the Appendix at the end of this paper.

The atmosphere in our simulations is configured based on Earth's atmosphere, with gravity ($g$) of 9.81$~\mathrm{m\, s^{-2}}$ and dry atmosphere gas constant ($R_d$) of 287$~\mathrm{J\, K^{-1}\, kg^{-1}}$. The domain simulated is 2D along the equator, which means there is no Coriolis force. The size of the domain is $35~\mathrm{km}$ in height and $40,000~\mathrm{km}$ in the zonal direction, representing the global atmosphere over the Equator of an Earth-size planet. The horizontal resolution is $2~\mathrm{km}$, and the vertical resolution is $250~\mathrm{m}$ (uniform, a simple set). The CFL number is set to 0.6, which means a time step of about $2~\mathrm{s}$. Sub-grid entrainment and detrainment are not included. There is no realistic radiation transfer process for either gases or clouds in the model. The longwave radiation is simulated simply by imposing a fixed radiative cooling rate everywhere (by default). As the heating effect of ozone gas by absorbing ultraviolet radiation is not calculated in the model, we set a minimum stratospheric temperature ($T_{st}$) of $170~\mathrm{K}$ to represent the stratosphere: when the air temperature drops below the value of $T_{st}$, a Newtonian adjustment back to $T_{st}$ is employed. 

We use periodic boundary condition in zonal direction and reflecting boundary condition in vertical direction. A sponge layer is added within the top $5~\mathrm{km}$ of the model ($H_{top}$) to absorb unrealistic reflecting gravity waves from the model top. A simple friction layer of $1~\mathrm{km}$ ($H_{low}$) is added at the bottom of the model as a momentum dissipation to simulate the effect of boundary layer turbulence and surface friction. Both the sponge and dissipation layers share the idealized form of:
\begin{equation}
    \frac{\mathrm{d}u(z)}{\mathrm{d}t}=-\frac{u}{\tau_i}\sin^{1/2}\left(\frac{\pi}{2}\frac{H_i-\Delta z}{H_i}\right), \quad i=top,\, low,
    \label{eq:momentum_diss}
\end{equation}
with $\tau_{top}$ is 2$~\mathrm{hrs}$, and $\tau_{low}$ is also $2~\mathrm{hrs}$ by default. $\Delta z$ is the distance between the altitude ($z$) and the top or lower boundary. We performed one series of sensitivity tests of the lower boundary friction parameter $\tau_{low}$ in the Appendix.


\begin{deluxetable*}{clc}[ht]
\tablenum{1}
\tablecaption{Experimental Designs Using the Cloud-Resolving Model SNAP}\label{tab:settings}
\tablewidth{0pt}
\tablehead{
\colhead{} & \colhead{Description} & \colhead{Value}
}
\startdata
$L$            & domain length (km)                                      & 40,000                \\
$H$            & domain height (km)                                      & 35               \\
$\Delta x$     & horizontal resolution (km)                              & 2                     \\
$\Delta z$     & vertical resolution (m)                                 & 250                   \\
$T_s^{max}$  & maximum surface temperature in the control run (K)          & 310 \\
$\Delta T_s$ & day-night surface temperature contrast in the control run (K) & 50 \\
$R_c$ & radiative cooling rate in the control run ($\mathrm{K\, day^{-1}}$) & $-1.89$\\
$\tau_{low}$     & surface friction time scale (hrs) & 2 \\
$k_2$  & conversion rate (cloud to precipitation) ($\mathrm{s^{-1}}$) & $10^{-4}$ \\
$k_4$  & re-evaporation rate (rain droplets to vapor) ($\mathrm{s^{-1}}$) & $10^{-2}$ \\
Group 1 & uniformly changing surface temperature $T_s$ (K)                 & 230\_280$^{a}$, 240\_290, 250\_300, 260\_310 \\
Group 2 & changing surface temperature contrast $\Delta T_s$ (K) & 260\_310, 275\_310, 290\_310, 305\_310 \\
Group 3 & changing radiative cooling rate $R_c$ ($\mathrm{K\, day^{-1}}$) & $-0.945$, $-1.89$, $-3.78$
\enddata
\tablecomments{a. The values are the minimum and maximum values of surface temperature, respectively (Figure~\ref{fig:SST}).}
\end{deluxetable*}


\begin{figure}[t]
    \centering
    \includegraphics[width=0.9\textwidth]{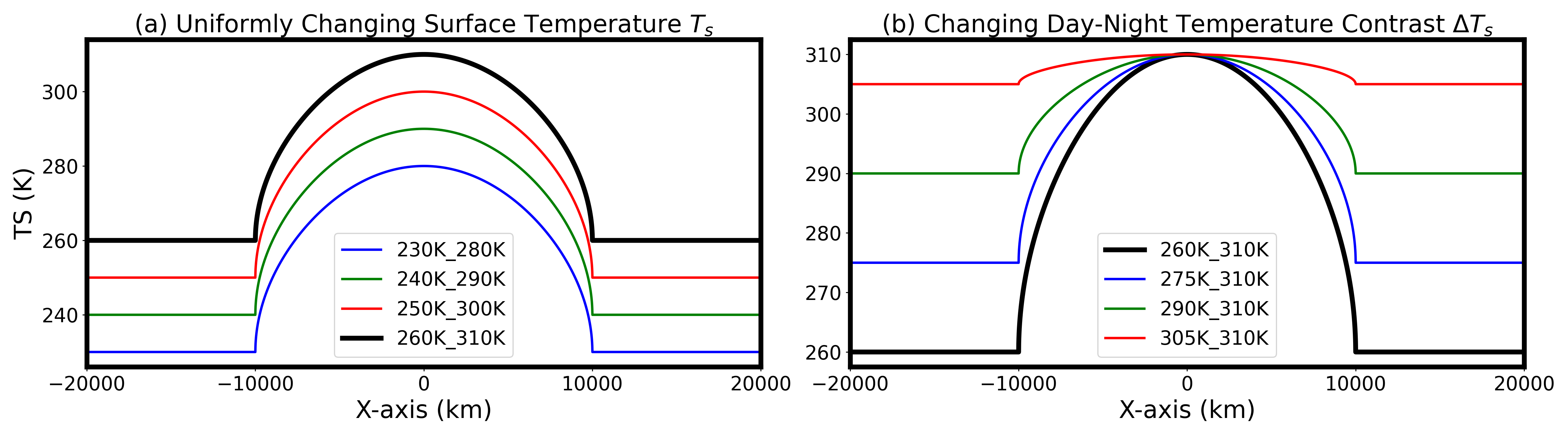}
    \caption{The surface temperature ($T_S$) distribution in the CRM experiments: (a) the group of uniformly changing $T_s$ by intervals of $10~\mathrm{K}$; (b) changing day-night $T_s$ contrast by intervals of $15~\mathrm{K}$ while fixing the maximum $T_s$ (310~K). The black line is the control experiment. The substellar point is at $x=0$, and the terminator line is at $x=\pm10,000~\mathrm{km}$. The legends show the minimum and maximum values of $T_s$.}
    \label{fig:SST}
\end{figure}

In our simulations, surface temperatures ($T_s$) are fixed (see Figure~\ref{fig:SST}), and their values are set roughly according to the simulated equatorial results of GCMs coupled to a slab ocean: on the dayside where $|x|<L/4$, $T_s(x) = T_s^{min} + \Delta T_s\cos^{1/2}\left(2\pi x/L\right)$, and on the nightside where $|x|\leq L/4$, $T_s(x)=T_s^{min}$. Here $L$ is the domain width, and $x$ varies from $-L/2$ to $L/2$. The power of 1/2 is used due to the weak temperature gradient in the substellar area and the strong temperature gradient near the western and eastern terminators. A plot of $T_s$ in different experiments is shown in Figure \ref{fig:SST}. In the control experiment, $T_s^{min}$ is 260 K, and $\Delta T_s$ is 50 K.

In calculating the surface fluxes of sensible heat and latent heat, the common treatments are discussed in \cite{Pierrehumbert_2010}. But since the model employed in this study is originally developed for simulating the atmosphere of gas giants, it does not have a detailed surface scheme yet. In the model, the surface fluxes of sensible heat and latent heat are simplified calculated using a Newtonian adjustment. The surface is adjusted to the specified surface temperatures and the saturated water mixing ratio at the corresponding surface temperatures within the time scale of $\tau_{sf}$. The value of $\tau_{sf}$ is two hours. These treatments capture the essential features that sensible and latent fluxes are proportional to the temperature and moisture differences between near-surface air and the surface.

We performed three groups of experiments to examine the responses of convection and clouds to different environmental settings, including varying the surface temperature everywhere uniformly, reducing the day-night surface temperature contrast, and changing the radiative cooling rate. Table \ref{tab:settings} summarises the experimental designs of the control experiment and the three different groups of experiments. Each experiment is run for different lengths of time to reach equilibrium, ranging from 30 to 100 Earth days. Then, an additional 20 Earth days of simulation is run to obtain the data for calculating the mean equilibrium states shown below.

\textcolor{black}{The radiative cooling rate, included in the diabatic heating term $\dot{Q}$ of Equation (6), is actually the net effect of radiation heating in stellar radiation and radiation cooling in thermal radiation. Since the CRM we used does not compute detailed radiation transfer processes, the prescribed uniform radiative cooling rate is an idealized setup for simplicity. The spatial distribution of radiative heating rate in real world or in 3D GCM experiments is non-uniform spatially (figures not shown). Overall, the stellar radiation has a warming effect on the dayside (similar to that of latent heat release), and the thermal radiation has a cooling effect on the nightside. In global and vertically-integrated mean, the net radiation heating rate should be negative, because it is balanced by the positive latent heat release and the mean sensible heat flux from the surface to the atmosphere (see Chapter 6 in \cite{hartmann2015global}). Therefore, in our 2D CRM simulations, we choose a net radiative cooling, rather than a net radiative heating. Moreover, to test the behavior of the atmospheric circulation and clouds in a non-uniform radiative cooling rate, we add one new experiment with a prescribed field of radiation warming (positive) on the dayside and meanwhile radiative cooling (negative) on the nightside. The result is shown in the Discussion section \ref{sec:discussion}.} \\


\subsection{3D GCM Simulations} \label{subsec:GCM_method}

The GCM we employed is the 3D Community Atmospheric Model version 3, called CAM3 \citep{collins2004description}, which is developed by the National Center for Atmospheric Research to simulate the climate of Earth. The model solves the primitive equations for atmospheric motion with a spectral Eulerian core on a rotating sphere. The radiative transfer model in CAM3 is based on \cite{ramanathan1986nonisothermal} and \cite{briegleb1992delta} with the updates described in \cite{collins2004description}. The radiation calculation includes the effects of water vapor, greenhouse gases, and clouds. The model uses subgrid-scale parameterizations to model deep and shallow/middle convection, condensation, precipitation, and evaporation of precipitating droplets  \citep{sundqvist1988parameterization, zhang1995sensitivity, rasch1998comparison, zhang2003modified}. Clouds are parameterized into three categories: marine stratus clouds, layered clouds, and convective clouds. Cloud fraction and the associated optical properties are evaluated via a diagnostic method. The diagnosis is a generalization of the scheme introduced by \cite{slingo1987development}, with variations described in \cite{hack1993description}, \cite{kiehl1998national}, and \cite{rasch1998comparison}. The model has 26 vertical levels from the surface to the middle stratosphere ($\approx$3 hPa) at the horizontal resolution of T42, which is $2.8^\circ$ in latitude by 2.8$^\circ$ in longitude.

In order to compare the results of CRM with those of GCM, we did \textcolor{black}{two groups of GCM experiments with no effect of planetary rotation and three groups with rotation effects included. For the experiments without rotation: the first group is for uniformly changing the surface temperature everywhere, and the second group is for reducing the day-night temperature contrast. We exclude the effect of rotation by setting a very long rotation period, $10^{30}$ seconds. In these two groups, the stellar flux is fixed (1200~W\,m$^{-2}$), and the surface temperatures are specified, as shown in subsection \ref{subsec:GCM_results} below. These no-rotation experiments are meant to fill in the gap between idealized 2D CRM experiments without rotation and 3D global GCM experiments with rotation.}

\textcolor{black}{For the experiments with rotation, we performed two same groups of experiments (uniformly changing the surface temperature and reducing the day-night temperature contrast, respectively) as the no-rotation experiments but with a rotation period of 30 Earth days.} We also performed another with-rotation group of experiments for varying stellar flux. In this group, the model is coupled with a slab ocean, which is a 50 m mixed layer, immobile ocean, with ocean heat transport specified to be zero everywhere. A series of stellar fluxes are employed in this group, 1200, 1400, 1600, 1800, 2000, and 2200 W\,m$^{-2}$. 

All the experiments are set to be 1:1 tidally locked, which means the rotation period is equal to the orbital period. The stellar temperature is set to be 3400 K. The atmosphere is Earth-like, including $\mathrm{N_2}$ and $\mathrm{H_2O}$. CO$_2$ concentration is 300 parts per million by volume (ppmv), CH$_4$ concentration is 0.8 ppmv, and N$_2$O is 0.27 ppmv. There is neither aerosol nor ozone. The model is coupled to a thermodynamic sea ice model, in which sea ice flows are not considered. In the experiments of \textcolor{black}{specified surface temperature (including the two groups without rotation and the first and second groups with rotation}), each case is integrated for 30 Earth years and reaches a steady state after about 20 years. In the  \textcolor{black}{experiments coupled with a slab ocean}, each case is integrated for about 60 Earth years and reaches a steady state after about 40 years. Averages over the last 5 years are used for our analyses below.


\section{Results} \label{sec:results}

\subsection{Results of the 2D CRM simulations}\label{subsec:CRM_results}


\begin{figure}[t]
    \centering
    \includegraphics[width=\textwidth]{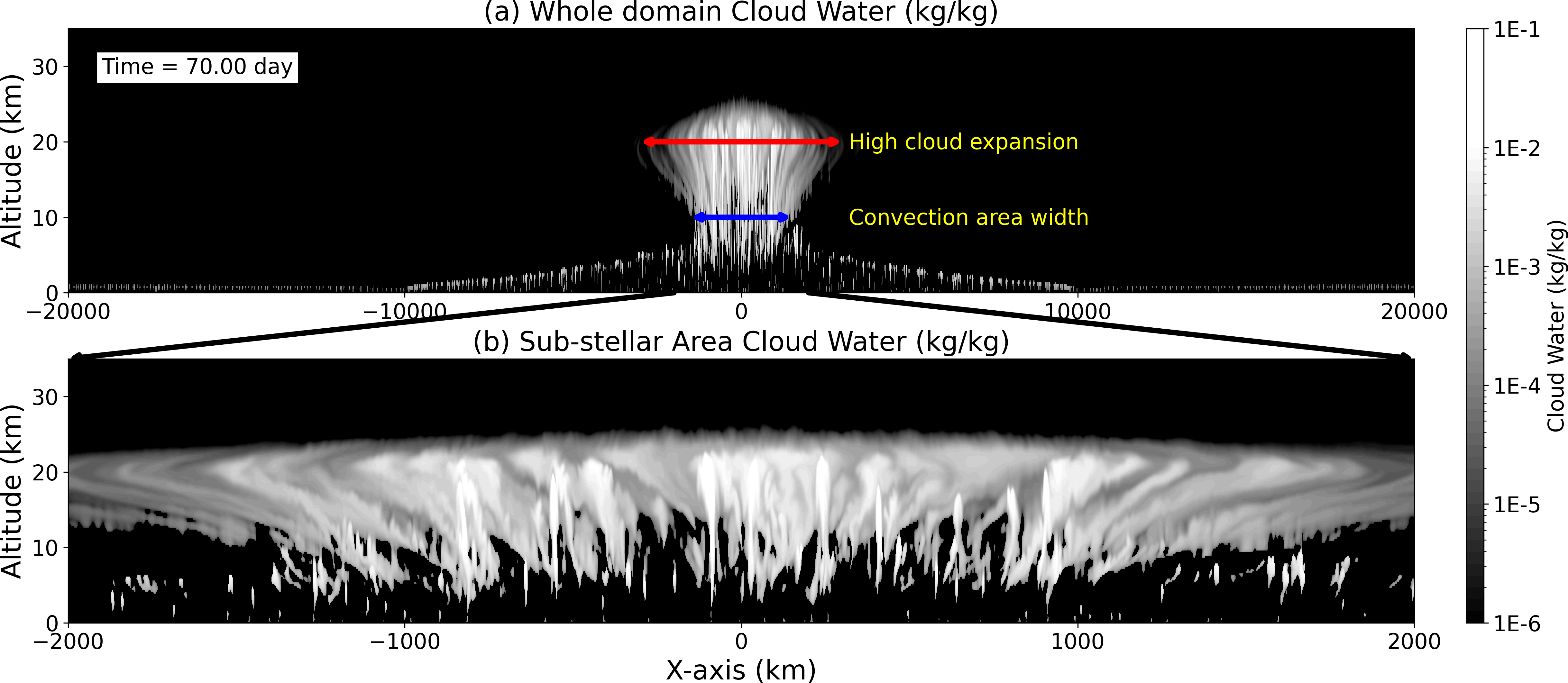}
    \caption{A snapshot of cloud water condensation in the control experiment: (a) the whole domain; (b) the substellar area. The substellar point is at $x=0$, and the terminator lines are at $x=\pm10,000~\mathrm{km}$. In (a), the width of the ascending branch of the large-scale overturning circulation (or roughly the width of the deep convective cloud area) and the width of high-level anvil clouds are marked with blue arrow and red arrow, respectively. In the control experiment, the surface temperature follows the black line shown in Figure~\ref{fig:SST}, and the radiative cooling rate is $-1.89~\mathrm{K\,day^{-1}}$ (a constant everywhere). An animation of the cloud water condensation is available in the online Journal. The animation follows the experiment for one model day (69-70).}
    \label{fig:snapshot}
\end{figure}


\begin{figure}[t]
    \centering
    \includegraphics[width=0.9\textwidth]{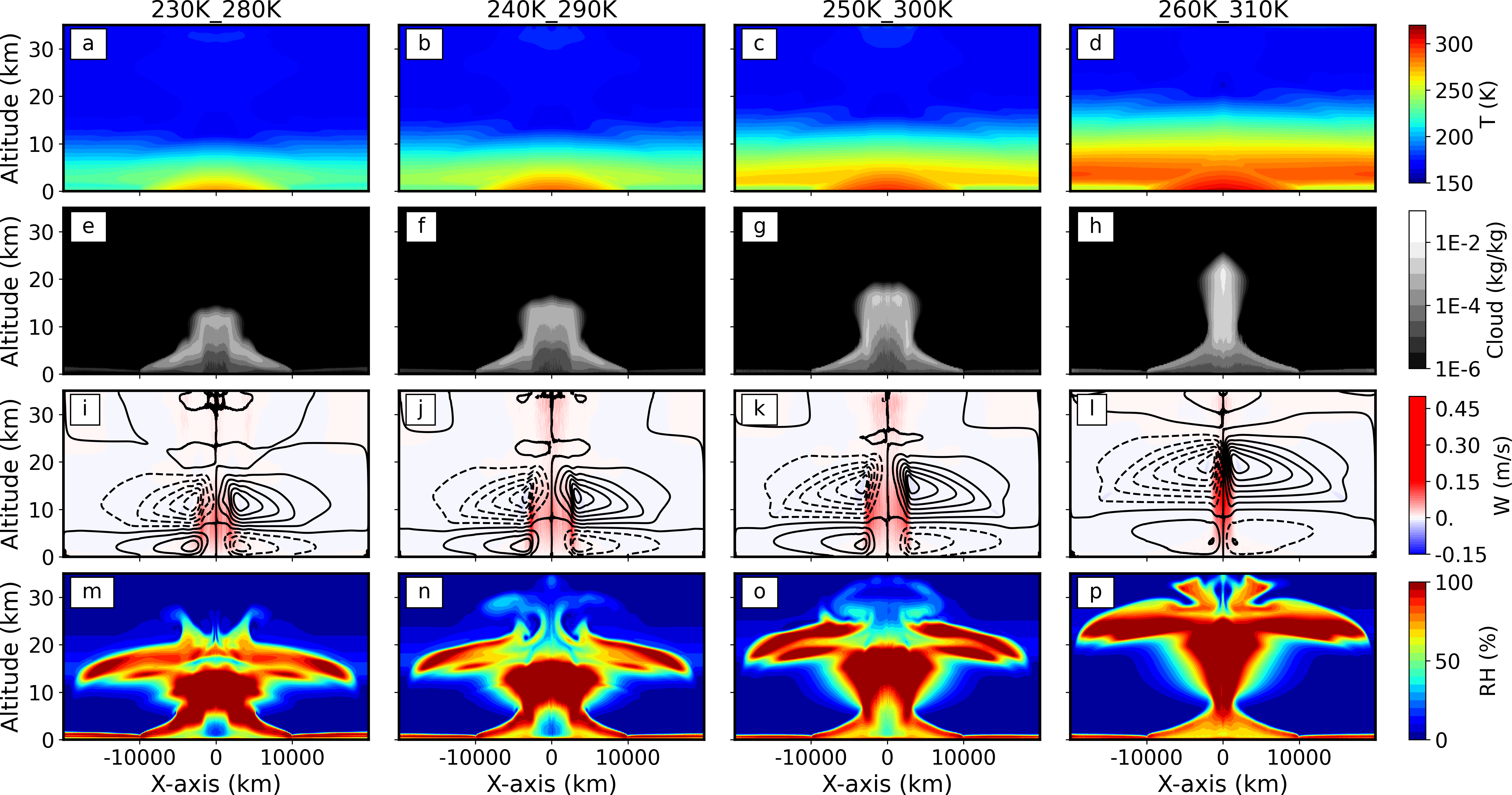}
    \caption{Responses of the atmosphere to uniform surface temperature changes. (a-d) air temperature (K), (e-h) cloud water concentration ($\mathrm{kg\, kg^{-1}}$) in logarithmic color scale, (i-l) vertical velocity ($\mathrm{m\,s^{-1}}$, color shaded) and horizontal velocity (contours with intervals of $5~\mathrm{m\, s^{-1}}$ and negative lines dashed), and (m-p) relative humidity (\%). The word in the top of each column, such as 230K\_280K, represents the minimum and maximum surface temperatures, respectively (see Figure~\ref{fig:SST}(a)).}
    \label{fig:group_1_state}
\end{figure}

\begin{figure}
    \centering
    \includegraphics[width=0.6\textwidth]{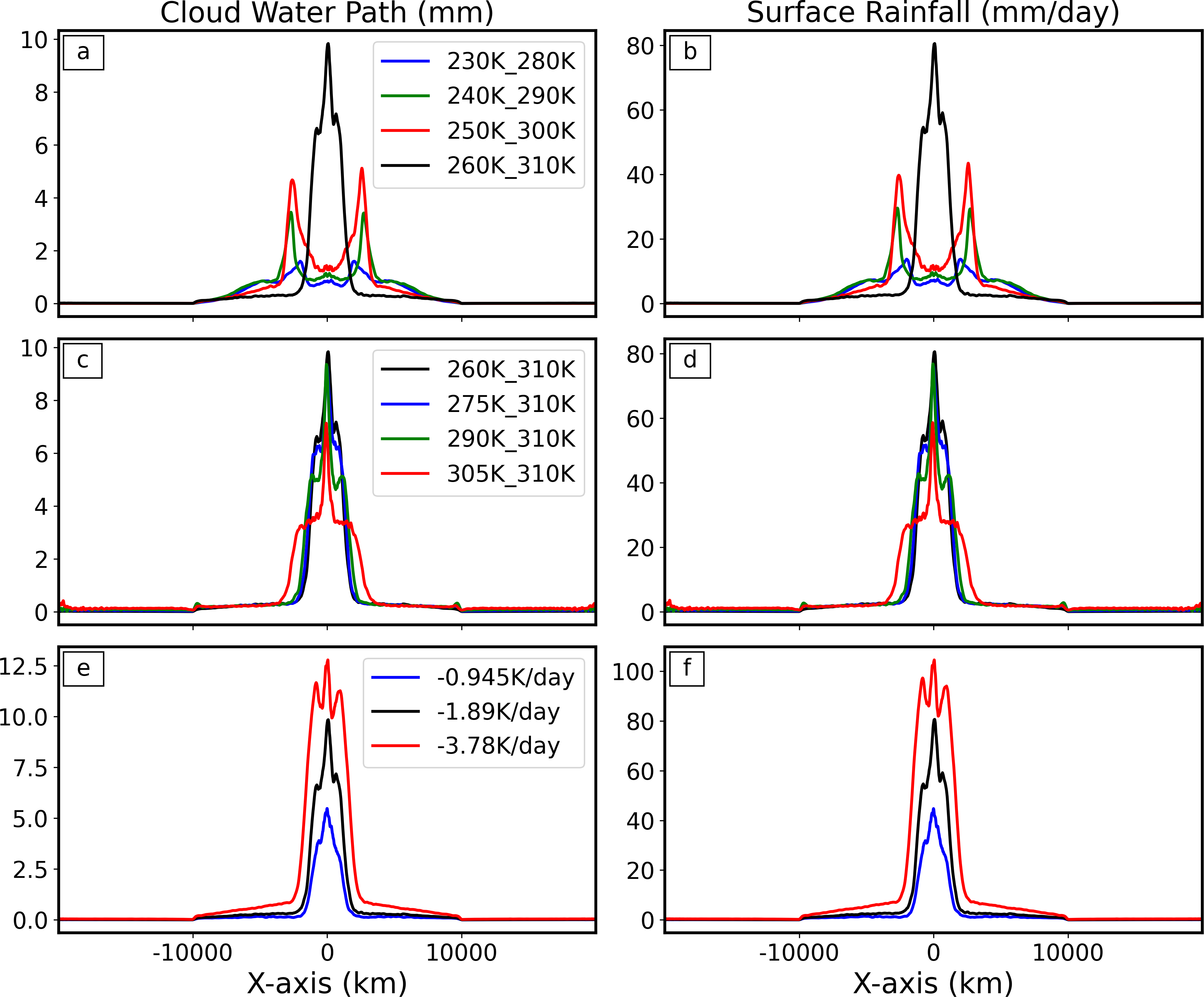}
    \caption{Effects of different environmental settings on the vertically integrated cloud water path (left column) and the surface rainfall rate (right column). Panels (a-b), (c-d), and (e-f) are for the three groups of experiments, respectively (see Table~\ref{tab:settings}).}
    \label{fig:state_1D}
\end{figure}

\begin{figure}[t]
    \centering
    \includegraphics[width=0.9\textwidth]{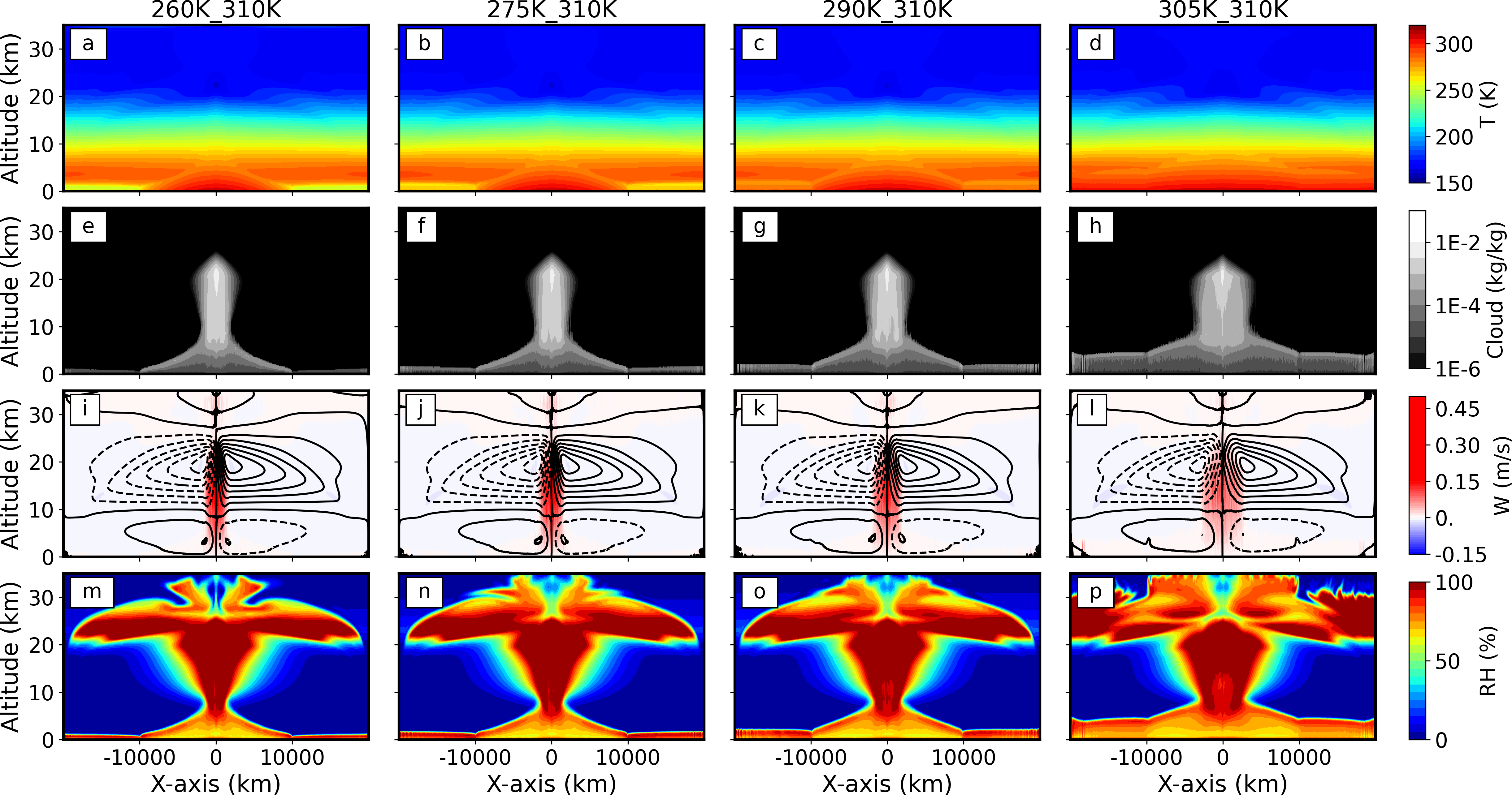}
    \caption{Same as Figure \ref{fig:group_1_state} but for the second group of experiments: changing day-night surface temperature contrast (see Figure~\ref{fig:SST}(b)). The maximum surface temperature is 310~K in all four experiments, but the night-side surface temperature is 260, 275, 290, and 305~K, respectively.}
    \label{fig:group_2_state}
\end{figure}

\begin{figure}
    \centering
    \includegraphics[width=0.9\textwidth]{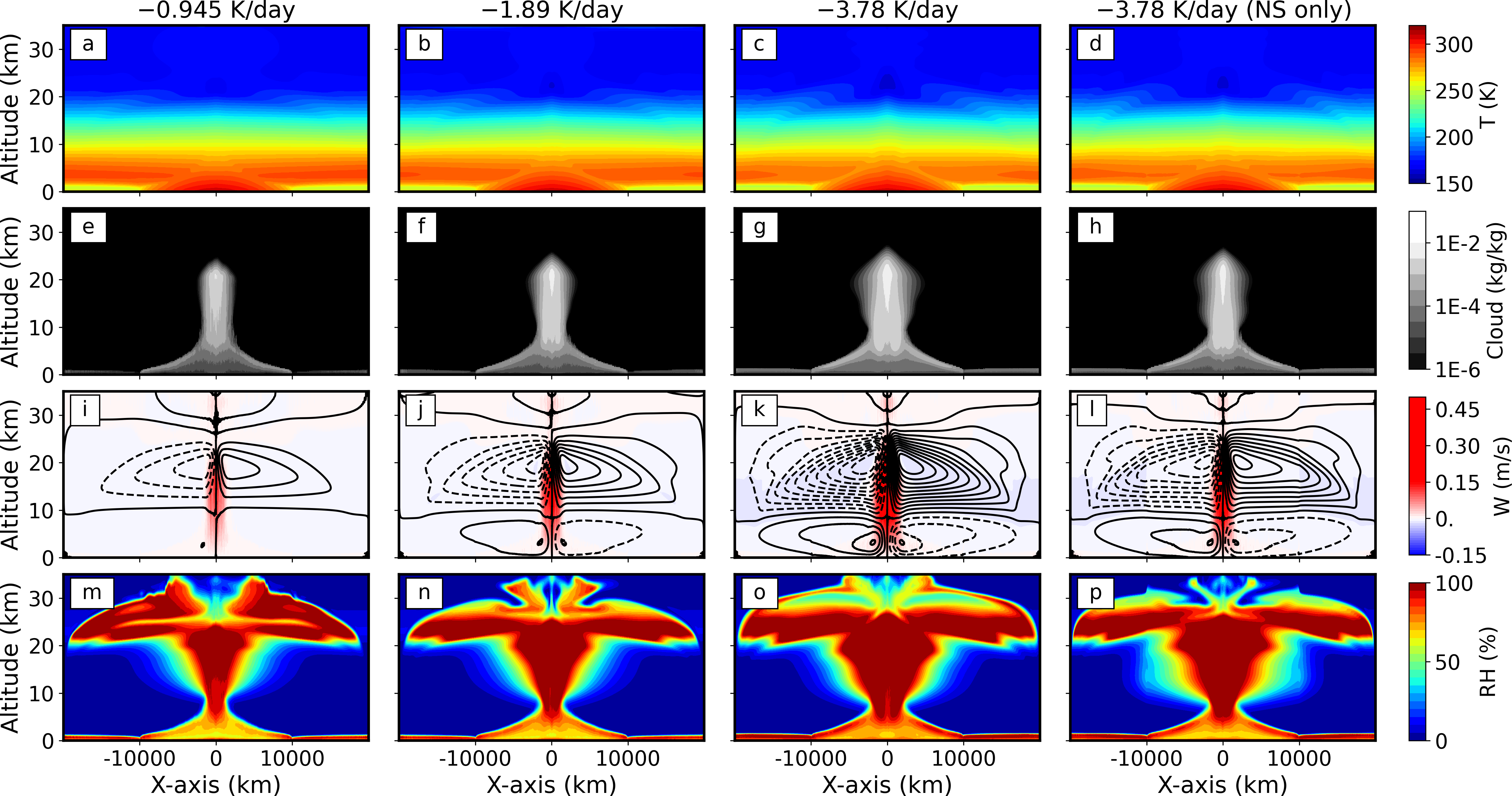}
    \caption{Same as Figure \ref{fig:group_1_state} but for the third group of experiments: changing the radiative cooling rate. From left to right, the radiative cooling rate is $-0.945$, $-1.89$ (control value), and $-3.78~\mathrm{K\, day^{-1}}$, respectively. In the fourth column, only the night-side (NS) radiative cooling rate is doubled (from $-1.89$ to $-3.78~\mathrm{K\, day^{-1}}$) with the day-side radiative cooling rate, $-1.89~\mathrm{K\, day^{-1}}$, unchanged.}
    \label{fig:group_3_state}
\end{figure}

In the control experiment, exerted by the centrally-peaked surface temperature pattern and spatially-uniform radiative cooling pattern, the most dominant character of the simulation results is the deep convection clouds in the substellar area (Figure \ref{fig:snapshot}; for \textcolor{black}{a clearer view of} the clouds, please see the online animation). The deep convection clouds are coupled with a global-scale Walker-like circulation that covers the whole troposphere (Figure \ref{fig:group_1_state}(l)). The airflow convergences in the lower atmosphere, ascends in the substellar area, reaches its top at the substellar tropopause (about $25~\mathrm{km}$), flows out at the altitude between 10 and $25~\mathrm{km}$, and then descends in the region outside of the substellar area. The area of the ascending branch is much narrower than that of the descending branch. The maximum speed of the convergent winds in the lower atmosphere is about $10~\mathrm{m\, s^{-1}}$, and that of the divergent winds in the upper troposphere is about $30~\mathrm{m\, s^{-1}}$. The surface convergence/divergence is approximately symmetric about the substellar point (i.e., the location of the maximum surface temperature). In the descending branch, the low-level atmosphere is very stable as the near-surface temperature is lower than that of the bottom of the free atmosphere. 



The clouds can develop to the upper troposphere in the ascending branch of the circulation (blue arrow in Figure \ref{fig:snapshot}(a)). In the upper troposphere (10 to $25~\mathrm{km}$) where there is strong divergence, the clouds expand towards the nightside following the strong outflow and form the wider high-level anvil clouds (red arrow in Figure \ref{fig:snapshot}(a)) than the mid-altitude clouds. The pattern of moisture distribution also follows the circulation (Figure \ref{fig:group_1_state}(p)). The descending branch of the circulation forces an inversion layer below $5~\mathrm{km}$ in the non-convective area (Figure \ref{fig:group_1_state}(d)), which traps moisture below it and prohibit\textcolor{black}{s} the night-side clouds from growing upwards (Figure \ref{fig:group_1_state}(h)).
The characteristics of these results are similar to previous studies of equatorial-Pacific Walker circulation \citep{bretherton2002simple, Liu2008explicitly, su2008observed, wofsy2012cloud}, but our experiments hold a temperature contrast of tens of degrees between the warm region and the cool region, which is an order of magnitude larger than previous studies. These results are also similar to previous studies on tidally locked planets using GCMs \citep{yang2013stabilizing, kopparapu2016inner, kopparapu2017habitable, wolf2017assessing}.

In the cloud-resolving simulations, from transient cloud snapshots, it is clear to see the high spatial variability of convection clouds in the troposphere (Figure \ref{fig:snapshot}). Each convection plume occupies only several model grids horizontally (about 10 km), but neighboring plumes are separated by clear-sky air with a distance of several hundred kilometers. In the upper troposphere with strong divergence, the anvil clouds of convection plumes join together to form the more smoothly distributed integral high clouds at about $20~\mathrm{km}$.

Below, we show the responses of clouds and large-scale circulation to varying the surface temperature and atmospheric radiative cooling rate. The possible reasons for the trends of the responses will be addressed in the following subsections \ref{subsec:mech_body} and \ref{subsec:mech_anvil}.

In the first group of experiments with uniform surface warming, the height of the cloud top rises, the ascending area of the large-scale circulation shrinks, the width of the high-level anvil clouds also becomes narrower, but the strength of the ascending branch becomes stronger (Figure~\ref{fig:group_1_state}). These can be viewed from the vertically-integrated cloud water path (Figure \ref{fig:state_1D}(a)), surface rainfall rate (Figure \ref{fig:state_1D}(b)), cloud water concentration (Figure \ref{fig:group_1_state}(f-h)), vertical velocity and horizontal velocity (Figure \ref{fig:group_1_state}(j-l)), and relative humidity (Figure \ref{fig:group_1_state}(n-p)). Quantitatively, in the three experiments of 240K\_290K, 250K\_300K, and 260K\_310K (representing the minimum and maximum surface temperature in each experiment, respectively), the fractional widths of the ascending area (divided by the whole domain width) are 17.1\%, 15.8\%, and 7.1\%, and the fractional widths of high-level anvil clouds at their respective peak altitudes are 23.0\%, 22.9\%, and 17.0\%, respectively. The experiment of the coolest case of 230K\_280K is an exception in this group, with smaller widths of the ascending area (15.5\%) and of the high-level anvil clouds (18.6\%) (Figure \ref{fig:group_1_state}(e)). This exception may be due to that the surface is too cold and the convection is too weak, so that near-surface friction has a strong effect in weakening both the near-surface convergence and the substellar convection.

In the second group of experiments with decreasing day-night temperature contrast, the area of the ascending branch widens, and the width of the high-level anvil clouds also becomes wider, but the strength of the ascending/descending branches becomes weaker. These can be viewed from the vertically-integrated cloud water path (Figure \ref{fig:state_1D}(c)), surface rainfall rate (Figure \ref{fig:state_1D}(d)), cloud water concentration (Figure \ref{fig:group_2_state}(e-h)), vertical velocity and horizontal velocity (Figure \ref{fig:group_2_state}(i-l)), and relative humidity (Figure \ref{fig:group_2_state}(m-p)). In the four experiments of 260K\_310K, 275K\_310K, 290K\_310K, and 305K\_310K, the fractional widths of the ascending area are 8.7\%, 9.3\%, 10.6\%, and 15.1\%, and the fractional widths of high-level anvil clouds at their respective peak altitudes are 17.0\%, 17.4\%, 18.2\%, and 21.9\%, respectively.

In the third group of experiments, when enhancing the strength of the longwave radiative cooling, both the ascending area and the high-level anvil clouds become wider, and the strength of the global overturning circulation becomes stronger. These can be viewed from the vertically-integrated cloud water path (Figure \ref{fig:state_1D}(e)), surface rainfall rate (Figure \ref{fig:state_1D}(f)), cloud water concentration (Figure \ref{fig:group_3_state}(e-g)), vertical velocity and horizontal velocity (Figure \ref{fig:group_3_state}(i-k)), and relative humidity (Figure \ref{fig:group_3_state}(m-o)). As the radiative cooling rate strengthens from $-1.89~\mathrm{K\,day^{-1}}$ to $-3.78~\mathrm{K\,day^{-1}}$, the atmospheric temperature drops slightly in the upper troposphere (Figure \ref{fig:group_3_state}(a-c)). In the three experiments of $-0.945~\mathrm{K\,day^{-1}}$, $-1.89~\mathrm{K\,day^{-1}}$, and $-3.78~\mathrm{K\,day^{-1}}$, the fractional widths of the ascending area are 8.6\%, 8.7\%, and 10.7\%, and the fractional widths of high-level anvil clouds at their respective peak altitudes are 13.1\%, 17.0\%, and 24.6\%, respectively (Figure~\ref{fig:group_3_state}(e-g)). In the added experiment of increasing the radiative cooling on the nightside only (from $-1.89~\mathrm{K\,day^{-1}}$ to $-3.78~\mathrm{K\,day^{-1}}$; the day-side radiative cooling is not changed; see the rightest column in Figure~\ref{fig:group_3_state}), the strength of the overturning circulation becomes stronger, and the cloud width become wider but with much smaller magnitudes, comparing to the experiment of increasing the radiative cooling of both nightside and dayside. This result suggests that both dayside and nightside radiative cooling can influence the convection and clouds.\\


\subsection{What Determines the Width and Strength of the Convection and Large-scale Circulation?}\label{subsec:mech_body}

\begin{figure}[t]
    \centering
    \includegraphics[width=0.6\textwidth]{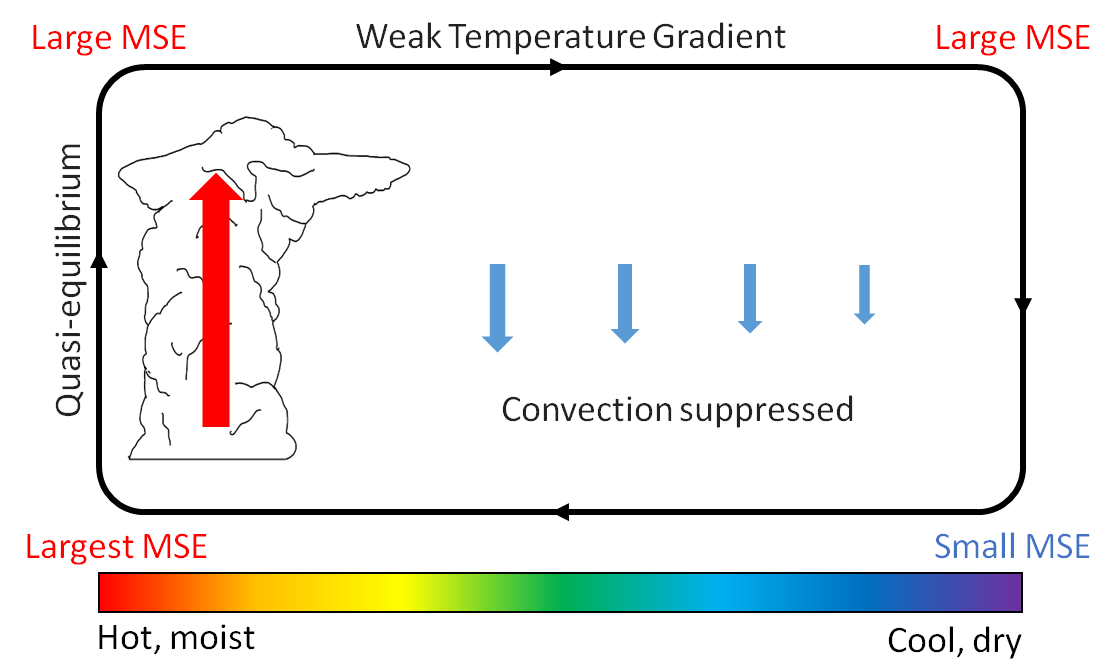}
    \caption{Conceptual illustration of the convection and large-scale atmospheric circulation. The warm and moist air in the substellar lower atmosphere results in the highest moist static energy (MSE) there, which is much larger than that of the night-side near-surface atmosphere, forming a large zonal MSE contrast. Following convective quasi-equilibrium, MSE in the upper troposphere of the substellar area is also high, close to that at the cloud base. In the free troposphere, the weak temperature gradient (WTG) approximation causes the air temperature in the free troposphere to be nearly uniform at each altitude. Therefore, on the nightside, MSE in the free atmosphere is much larger than MSE near the surface, suppressing convection activity there.}
    \label{fig:concept}
\end{figure}

\begin{figure}[t]
    \centering
    \includegraphics[width=\textwidth]{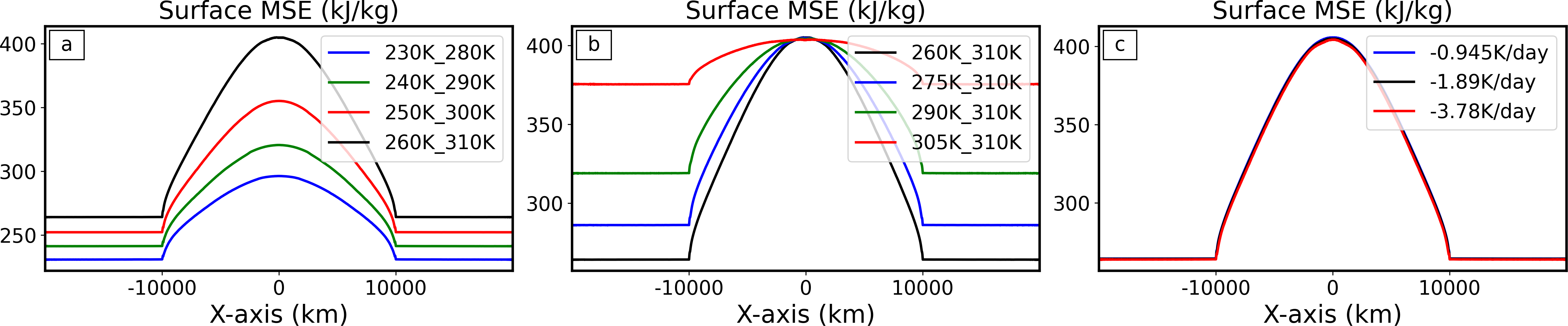}
    \caption{Surface MSE in the three groups of experiments: (a) uniformly changing the surface temperature, (b) changing the day-night surface temperature contrast, and (c) changing the radiative cooling rate.}
    \label{fig:state_1D_MSE}
\end{figure}

\begin{figure}[t]
    \centering
    \includegraphics[width=0.9\textwidth]{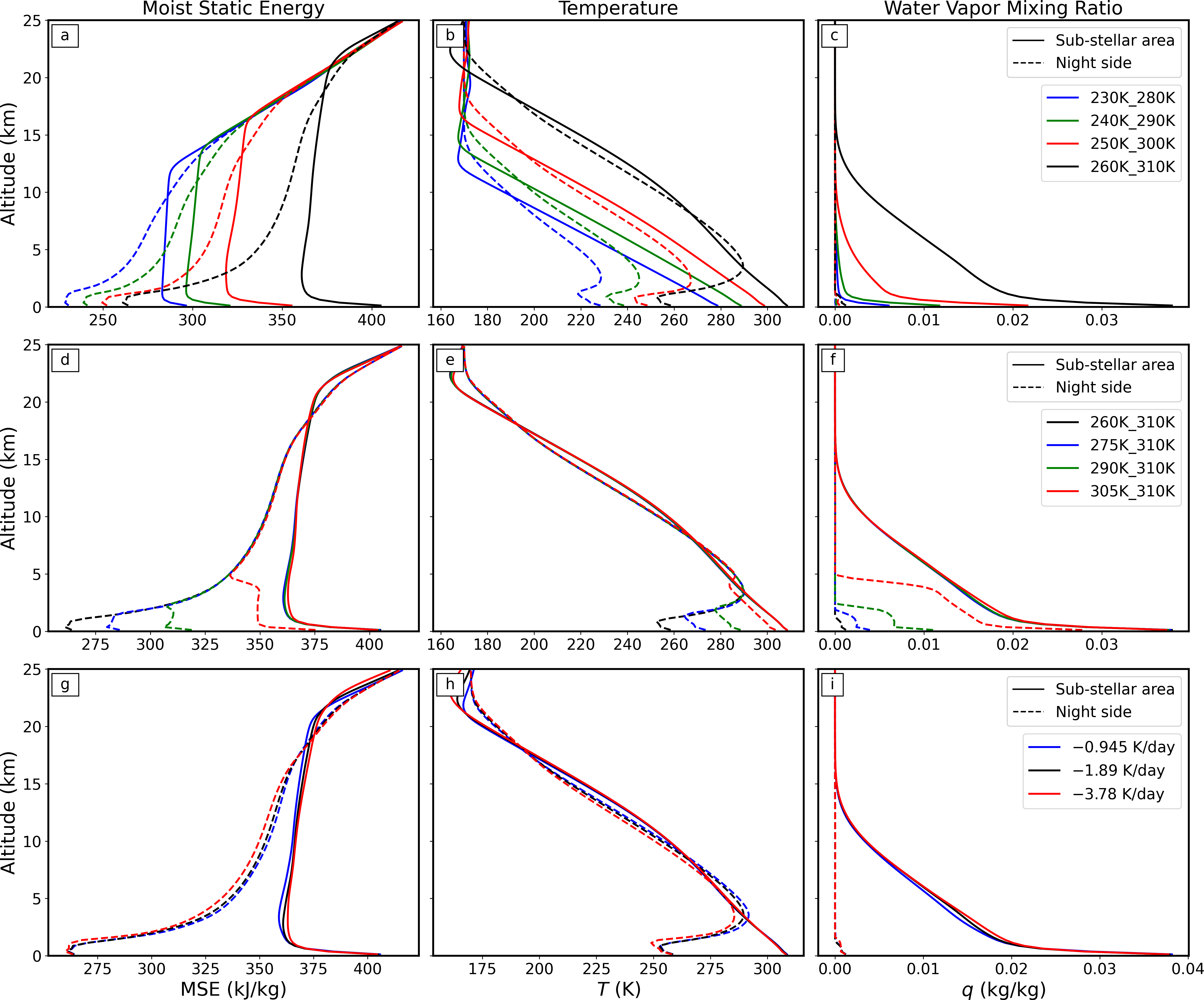}
    \caption{Vertical profiles of the three groups of experiments. The left column is MSE profiles, the center column is air temperature profiles, and the right column is water vapor mixing ratio. Solid lines are for the area mean over the substellar region (representing the ascending branch of the large-scale atmospheric overturning circulation), and dashed lines are for the area mean over the nightside (representing the descending branch). (a-c) uniformly changing the surface temperature, (d-f) changing the day-night surface temperature contrast, and (g-i) changing the radiative cooling rate.}
    \label{fig:MSE}
\end{figure}

In subsection \ref{subsec:CRM_results}, we have found that the widths of the ascending area and of the mid-altitude clouds decrease with uniform surface warming but increase under reduced day-night surface temperature contrast or under enhanced longwave radiative cooling. And, the strengths of the convection and large-scale circulation become stronger under uniform surface warming or enhanced longwave radiative cooling but become weaker when the day-night surface temperature contrast decreases. An analysis of moist static energy (MSE) can roughly explain these trends. MSE is a thermodynamic variable that describes the state of a moist air parcel, and it is the sum of the parcel's enthalpy ($c_pT$), the geopotential energy ($\Phi$), and the latent energy ($L_vq$). The MSE is written as:
\begin{equation}
    MSE=c_pT+\Phi+L_vq,
\end{equation}
where $q$ is the water vapor mixing ratio. The conceptual diagram for the process of how MSE constrains the convection and large-scale circulation on tidally locked habitable planets is shown in Figure \ref{fig:concept}. The zonal distribution of surface MSE in the three groups of experiments is shown in Figure~\ref{fig:state_1D_MSE}. The vertical profiles of MSE and its temperature and vapor terms in the three groups are shown in Figure \ref{fig:MSE}, in which the substellar profiles represent the ascending area, and the night-side profiles represent the descending area.

MSE is a useful variable because it is approximately conserved during adiabatic ascent or descent with phase changes of water between liquid and vapor \citep{wallace2006atmospheric}. So, the vertical profile of MSE can be used to inform the strength of convection. In the troposphere, a constant MSE profile represents a thermodynamically neutral atmospheric stratification. When the MSE in the free atmosphere is lower than that near the surface, convection tends to develop and release the instability of the atmosphere. When the vertical structure of MSE is inverted with MSE near the surface being lower than that of the free atmosphere, the atmosphere tends to be stable.

Because the surface atmosphere in the substellar area is much warmer than that on the nightside, the day-side near-surface MSE is much greater than that of the nightside (Figure \ref{fig:state_1D_MSE}). In the warm substellar area, the atmosphere is in a quasi-equilibrium state, for which any existing buoyancy in the atmosphere can be adjusted by moist convection, and the saturated MSE in the free troposphere equals the subcloud MSE where the air parcel rises from. This adjustment process takes a very short time compared with external forcing or large-scale mean circulation because convection in the atmosphere is a fast process. As a result, the atmosphere almost stays in a quasi-equilibrium state over the substellar region, and MSE in the upper troposphere is tightly connected to MSE near the surface \citep{arakawa1974interaction, marquet1993exergy, emanuel1994large, emanuel2007quasi, neelin2008rethinking}. Therefore, MSE in the free troposphere over the substellar region is also high, as shown in Figure~\ref{fig:MSE}(a, d, and g).

The horizontal temperature gradient in the upper troposphere is weak, satisfying the weak temperature gradient (WTG) approximation \citep{charney1963note, sobel2001weak}. This is because of the nonexistence of the horizontal Coriolis force along the equator, and thus the momentum and heat transports by gravity waves and large-scale advection in the free atmosphere are effective. Consequently, the high MSE in the substellar area can spread towards the nightside, resulting in almost uniformly high MSE in the upper troposphere of the whole domain. Therefore, on the nightside, there exists a large vertical inversion in MSE, with the value in the upper troposphere being much higher than that near the surface. The inversion area also extends across the terminator lines, occupying a large area of the dayside. This MSE inversion implies that the atmosphere is stable, and convection activity is prohibited there.

Following the illustration above, under the quasi-equilibrium approximation and the WTG approximation, the strength of the convection over the substellar region is constrained by the MSE difference between the near-surface and the free troposphere, and the fraction of the convective (as well as non-convective) area could be mainly determined by the zonal MSE gradient.

In the first group, the increased strengths of the convection and the large-scale mean circulation could be explained by the enhanced MSE difference between the near surface and the free troposphere over the substellar region (Figure~\ref{fig:MSE}(a)). The narrowing trend of the convection area could be explained by the relatively larger increase of MSE over the substellar region than that on the nightside (Figure~\ref{fig:state_1D_MSE}(a)). This is because when the surface temperature rises by the same magnitude, water vapor acts as an amplifier (Figure \ref{fig:MSE}(c)) since the saturated water vapor mixing ratio follows a strongly non-linear Clausius–Clapeyron relation with air temperature. The enhanced day-night MSE contrast results in a larger MSE inversion outside the convecting substellar area (see dashed lines in Figure~\ref{fig:MSE}(a)), thus suppressing the convection more effectively and leading to the narrowing trend of the ascending area.

In the second group, the widening trends of the convective cloud area and of the ascending branch of the large-scale circulation are due to the decrease of the MSE contrast between the substellar area and the rest region (Figures \ref{fig:state_1D_MSE}(b) \& \ref{fig:MSE}(d)). (1): MSE at the near surface of the nightside increases a lot, due to the direct effect of the warming on the nightside (Figure~\ref{fig:MSE}(e)); the increase of water vapor on the nightside also acts to amplify the night-side near-surface MSE and reduce the day-night MSE contrast (Figure~\ref{fig:MSE}(f)). (2): In the free troposphere, the thermodynamic profiles including MSE, temperature, and water vapor remain nearly the same for both dayside and nightside, when reducing the day-night contrast (Figure \ref{fig:MSE}(d-f)). This is because surface temperature over the substellar area nearly does not change, and because of the strong constraints of the quasi-equilibrium approximation and the WTG approximation. This means that the substellar surface temperature determines not only thermal properties near the surface but also those in the entire free troposphere. Strong convection links the free troposphere to the surface and the WTG approximation links the substellar area to the rest region in the free troposphere. (3): Due to the combination of (1) and (2), the vertical MSE inversion on the nightside decreases. Therefore, the suppression of convection is weakened, and the convection area expands towards the nightside. Moreover, the weakening of the atmospheric circulation can also be understood using the constrain of energy transport. Due to the constrain of the WTG approximation and the reduced day-night MSE contrast, less energy is required to be transported from the substellar region to the nightside. As the ascending branch becomes wider, the weakening of its strength is necessary.


In the third group, the widening trends of the convective clouds and the ascending branch of the large-scale circulation are constrained by horizontal energy transport. On the dayside, the main energy balance of the atmosphere is between latent heat release, radiative cooling, and horizontal energy output to the nightside \citep{Yang_abbot2014}. On the nightside, the main energy balance is between radiative cooling and horizontal energy input from the dayside. For global mean, the dominated balance is between latent heat release and radiative cooling (note that there is no explicit shortwave or longwave radiation in all the experiments of this study). 
When the radiative cooling is increased, on one hand, the magnitude of the latent heat release needs to increase in order to balance the increased radiative cooling. This implies increases in the strengths of convection and large-scale circulation, as found in Figure~\ref{fig:group_3_state}(e-g and i-k). On the other hand, the profiles of MSE, air temperature, and water vapor nearly do not change (Figure~\ref{fig:MSE}(g-i)). This implies that the atmospheric heat transport \textcolor{black}{effectively maintains} the weak temperature gradients. So, the strength of the large-scale circulation needs to increase to balance the enhanced radiative cooling on the nightside. Moreover, the enhanced radiative cooling directly drives stronger downwelling on the nightside and \textcolor{black}{strengthens} the global overturning circulation; in the following section, we will more exactly show how the radiative cooling directly influences the strength of the downwelling branch.

Furthermore, the latent heat release increases but with a smaller magnitude than that of the increase of radiative cooling rate. This can be found in Figure~\ref{fig:state_1D}(f): when the radiative cooling rate is doubled, the magnitude of the rainfall rate (equaling to the latent heat release over the constant of specific latent heat ($L_v$)) increases by only about 30\%. This is likely limited by the unchanged surface temperature in this group of experiments. The relatively smaller magnitude increase in latent heat release implies that the widths of the convective area and the ascending branch of the large-scale circulation are required to increase. \\


\subsection{What Determines the Width of the High-level Anvil Clouds?}\label{subsec:mech_anvil}

\begin{figure}[t]
    \centering
    \includegraphics[width=0.9\textwidth]{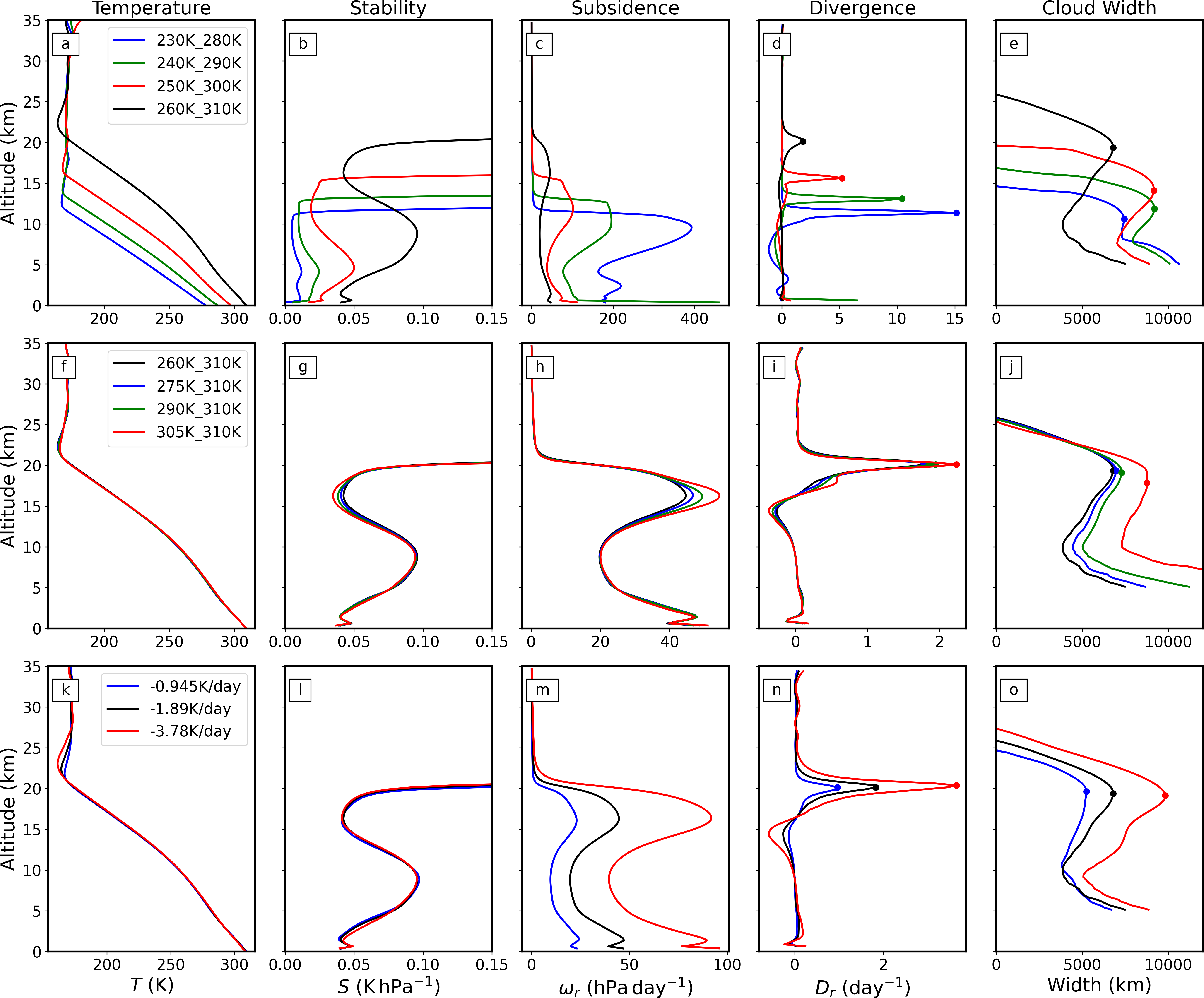}
    \caption{The mechanism that determines the width of the high-level anvil clouds. The first column is the vertical temperature profiles over the substellar area, the second column is the atmospheric stability, the third column is the radiation-driven vertical velocity (Equation (\ref{eq:Omega})), the fourth column is the corresponding horizontal divergence, and the fifth column is the cloud width simulated by the model SNAP. The cloud width is defined as the width of the area with temporally-averaged cloud water mixing ratio larger than $10^{-6}~\mathrm{kg\,kg^{-1}}$. The dots in the fourth and fifth columns mark the peak of the curves. The upper row is for the experiments of uniformly changing the surface temperature, the middle row is for the experiments of changing the day-night temperature contrast, and the lower row is for the experiments of changing the radiative cooling rate.}
    \label{fig:anvil}
\end{figure}

In subsection \ref{subsec:CRM_results}, we have shown that the width of the high-level anvil clouds shrinks as the surface temperature uniformly increases, but expands when reducing the day-night temperature contrast or increasing the radiative cooling rate. Below, we show that this can be qualitatively explained using radiation-driven divergence, generally following the idea of \cite{bony2016assessing} and \cite{cronin2017clouds}.
The high-level anvil clouds in the upper troposphere of the dayside are transported from the substellar convective area towards the nightside. This outflow is closely related to mass divergence. Based on mass conservation, horizontal divergence ($D_r$) \textcolor{black}{can be diagnosed by the clear-sky radiation-driven convergence} $\frac{\partial \omega_r}{\partial p}$, where $\omega_r$ is the \textcolor{black}{radiation-driven} vertical velocity in pressure coordinate, and $p$ is air pressure \citep{Vallis2017}. The value of $\omega_r$ could be obtained from the temperature equation in an equilibrium state. Under the WTG approximation, the horizontal temperature gradients can be ignored, so the vertical velocity ($\omega_r$) is approximately equal to: 
\begin{equation}
\omega_r \cong -\frac{Q_r}{S},
    \label{eq:Omega}
\end{equation}
where $Q_r$ is the radiative cooling rate, and $S$ is the atmospheric stability \citep{emanuel1994large,bony2016assessing}. The value of $S$ is defined as $-\frac{T}{\theta}\frac{\partial \theta}{\partial p}$, where $T$ is air temperature, and $\theta$ is potential temperature. Therefore, the width of the outflow (assuming to be proportional to the width of the high-level anvil clouds) is mainly determined by two factors: the strength of the radiative cooling and the atmospheric stability \citep{bony2016assessing}. The vertical profiles of $T$, $S$, $\omega_r$, $D_r$, and the width of the high-level anvil clouds are shown in Figure~\ref{fig:anvil}.

In the first group of experiments, the width of the high-level anvil clouds generally decreases with uniformly increasing the surface temperature (Figure~\ref{fig:anvil}(e)). This is mainly due to the increase of atmospheric stability ($S$, Figure~\ref{fig:anvil}(b)), and thereby the radiation-driven divergence becomes weaker (Figure~\ref{fig:anvil}(d)). The increase of $S$ is due to the fact that the warming of the atmosphere is larger than the warming of the surface, just following the moist adiabatic process. Note that the altitude of the anvil clouds exhibits a significant upward shift with surface warming. This is because the atmospheric convection reaches higher altitudes. In the second group, the width of the high-level anvil clouds increases when the day-night surface temperature contrast decreases (Figure~\ref{fig:anvil}(j)). It is due to the decrease of the atmospheric stability in the altitudes around 15--20 km (Figure~\ref{fig:anvil}(g)). In the third group, the width of the high-level anvil clouds increases with enhancing radiative cooling rate in the atmosphere (Figure~\ref{fig:anvil}(o)). This is due to the direct effect of increasing the value of $Q_r$ in Equation~(\ref{eq:Omega}), although the air temperature and the stability nearly do not change (Figure~\ref{fig:anvil}(k,l)). The latter is mainly determined by the surface temperature that is fixed in this group of experiments. \\


\subsection{Results of the 3D GCM Simulations}\label{subsec:GCM_results}

\begin{figure}[t]
    \centering
    \includegraphics[width=\textwidth]{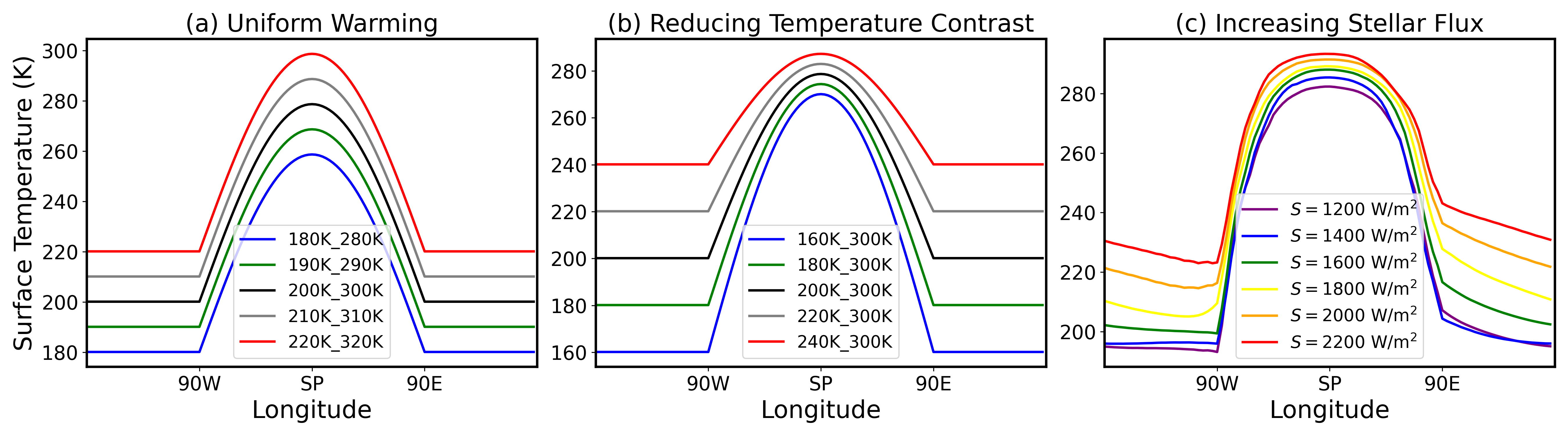}
    \caption{Surface temperature distribution in the GCM experiments. Panel (a) shows the specified surface temperatures in the groups \textcolor{black}{(with and without rotation)} of uniformly warming or cooling the surface by intervals of $10~\mathrm{K}$; panel (b) shows the specified surface temperatures in the groups \textcolor{black}{(with and without rotation)} of reducing the day-night contrast; and panel (c) shows the obtained surface temperatures in the simulations coupled to a 50-m slab ocean \textcolor{black}{(with rotation)}. These plots show the meridional-mean surface temperatures from $90^\circ$S and $90^\circ$N. \textcolor{black}{The label SP on the x-axis represents the substellar point.} Note that in panel (b) the meridional-mean surface temperatures in the substellar region also increase significantly, due to that this figure is for the mean from 90$^{\circ}$S to 90$^{\circ}$N, and both the day-night surface temperature contrast and the equator-pole surface temperature contrast decrease in this group of experiments.}
    \label{fig:SST_GCM}
\end{figure}

\begin{figure}[t]
    \centering
    \includegraphics[width=0.9\textwidth]{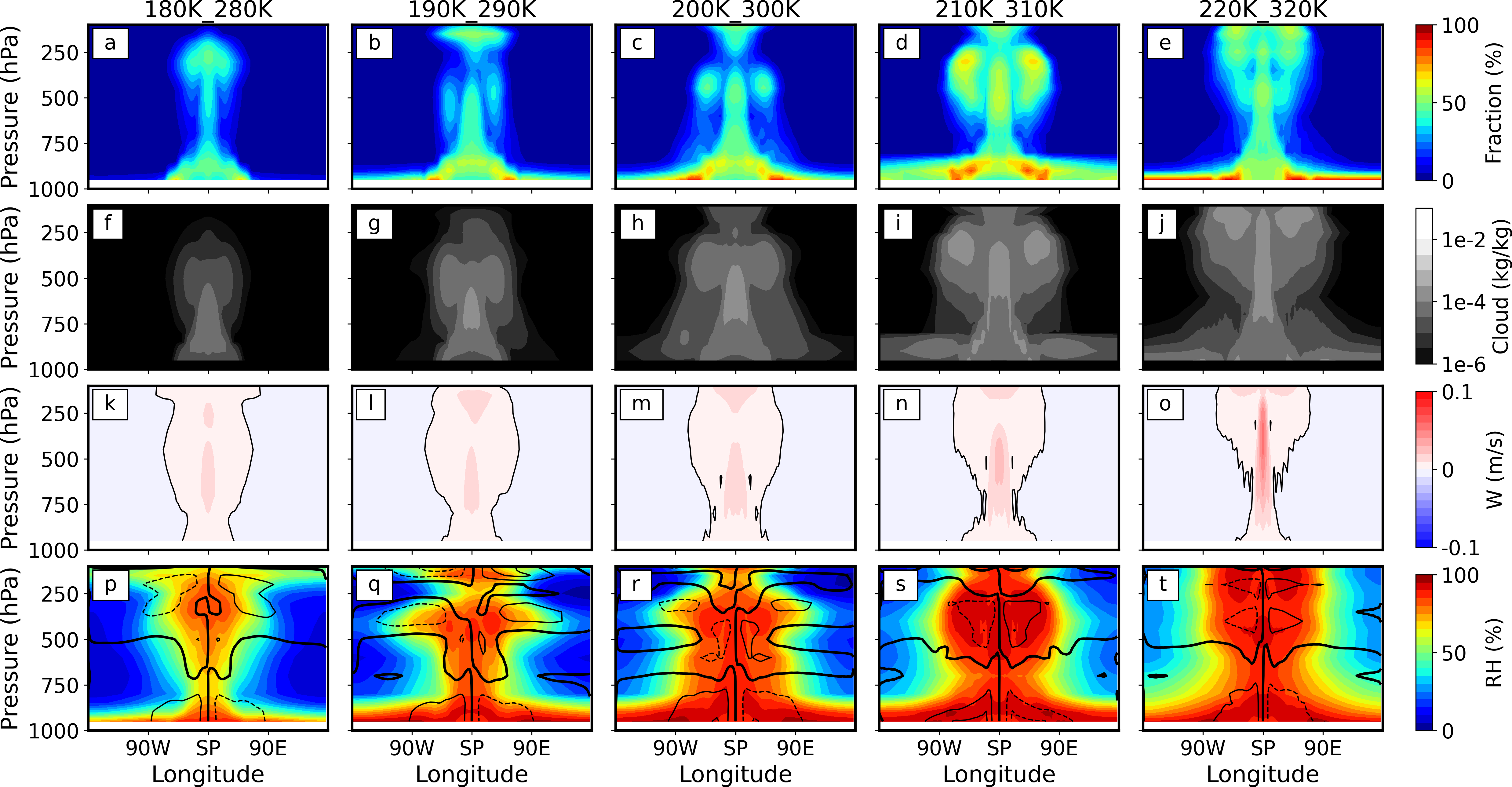}
    \caption{\textcolor{black}{Meridional-mean values from $90^\circ$S and $90^\circ$N for (a-e) cloud fraction, (f-j) cloud water concentration, (k-o) vertical velocity, and (p-t) relative humidity (color shaded) and zonal wind (contour lines with intervals of $5~\mathrm{m\,s^{-1}}$, bold zero-value line, and dashed negative lines) in the GCM experiments of uniformly changing the surface temperatures (Figure \ref{fig:SST_GCM}(a)) without rotation. Each column represents an experiment, and the minimum and maximum surface temperatures are denoted on the top of each column.}}
    \label{fig:GCM_uni_no_rotation}
\end{figure}

\begin{figure}[t]
    \centering
    \includegraphics[width=0.9\textwidth]{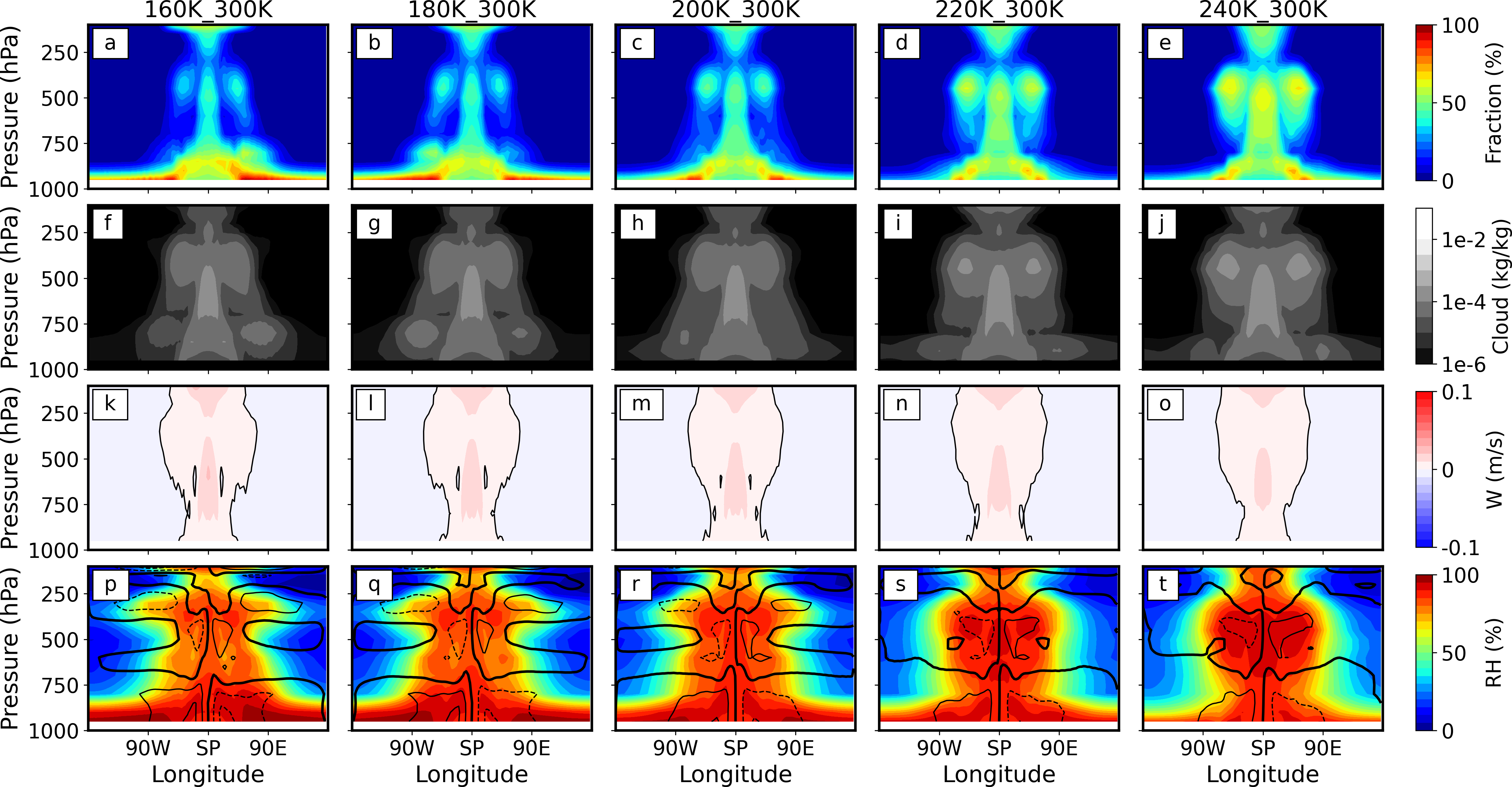}
    \caption{\textcolor{black}{Same as Figure \ref{fig:GCM_uni_no_rotation} but for the GCM experiments of changing the day-night surface temperature contrast (Figure \ref{fig:SST_GCM}(b)) without rotation.}}
    \label{fig:GCM_fixm_no_rotation}
\end{figure}

\begin{figure}[t]
    \centering
    \includegraphics[width=0.6\textwidth]{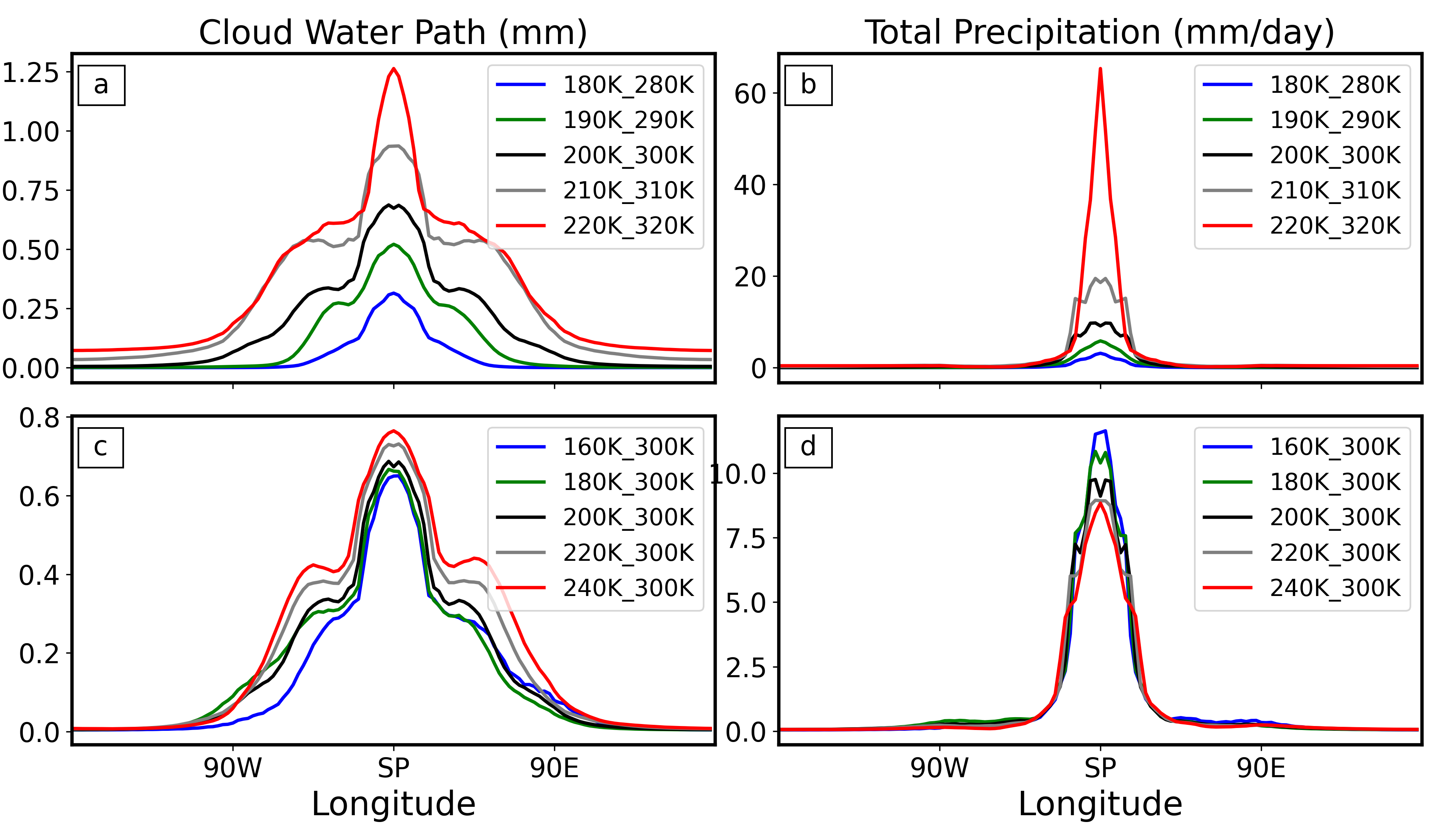}
    \caption{\textcolor{black}{Vertically integrated meridional-mean cloud water path (left column) and the surface rainfall rate (right column) under the two groups of GCM environmental configurations without rotation. Panels (a-b) are for uniformly increasing surface temperature, and panels (c-d) are for reducing day-night surface temperature contrast.}}
    \label{fig:GCM_1D_no_rotation}
\end{figure}

In order to \textcolor{black}{closely} compare the results of CRM simulations to the results of GCMs, we did \textcolor{black}{two} corresponding groups of experiments \textcolor{black}{without the effect of planetary rotation} using CAM3. In the first group, the surface temperatures are specified as shown in Figure~\ref{fig:SST_GCM}(a), and five cases are considered with uniformly varying the surface temperatures by 10~K in each case. This group is corresponding to the first group of CRM. In the second group, the surface temperature at the substellar point is fixed (300 K), while the night-side surface temperatures are changed by 20~K in each case, as shown in Figure~\ref{fig:SST_GCM}(b). This group is corresponding to the second group in the CRM experiments, although not 100\% similarity. The results of these \textcolor{black}{two} groups of experiments are shown in \textcolor{black}{Figures~\ref{fig:GCM_uni_no_rotation}, \ref{fig:GCM_fixm_no_rotation}, and \ref{fig:GCM_1D_no_rotation}}. Because realistic radiation transfer is included in the GCM, no experiment of fixed radiative cooling was performed. 

\textcolor{black}{In the group of experiments of uniformly increasing surface temperature without rotation, the cloud fraction and cloud water content over the substellar area increase significantly (Figure \ref{fig:GCM_uni_no_rotation}(a-j)). The ascending branch of the large-scale circulation strengthens and shrinks (Figure \ref{fig:GCM_uni_no_rotation}(k-o)), which is consistent with CRM results (Figure \ref{fig:group_1_state}(i-l)). The atmospheric relative humidity increases significantly (Figure \ref{fig:GCM_uni_no_rotation}(p-t)).}

\textcolor{black}{In the group of experiments of reducing day-night surface temperature contrast without rotation, the cloud fraction and cloud water path increase and the cloud extend becomes wider (Figure \ref{fig:GCM_fixm_no_rotation}(a-j)). The ascending branch of the large-scale circulation becomes weaker (Figure \ref{fig:GCM_fixm_no_rotation}(k-o)), which is consistent with CRM results (Figure \ref{fig:group_2_state}(i-l)). The atmosphere relative humidity shows little variance, increasing in the substellar upper atmosphere ($\approx500~\mathrm{hPa}$) and decreasing in the night-side low-level atmosphere below 850 hPa (Figure \ref{fig:GCM_fixm_no_rotation}(p-t)).}

\textcolor{black}{In terms of cloud water path and surface precipitation, the results in these two groups show good symmetry between the east and west of the substellar point (Figure \ref{fig:GCM_1D_no_rotation}), which is similar to the CRM results within which planetary rotation is also excluded (Figure \ref{fig:state_1D}). When uniformly increasing the surface temperature, the increases in peak values of cloud water path and precipitation (Figure \ref{fig:GCM_1D_no_rotation}(a,b)) are consistent with CRM results (Figure \ref{fig:state_1D}(a,b)), but the increase in cloud width is opposite to the narrowing cloud in CRM results. When reducing the day-night surface temperature contrast, the changes in both cloud water path and precipitation are small (Figure \ref{fig:GCM_1D_no_rotation}(c,d)), but the peak value of cloud water path increases with the reduction of temperature contrast, which is opposite to the CRM results (Figure \ref{fig:state_1D}(c,d)).}


\begin{figure}[t]
    \centering
    \includegraphics[width=0.9\textwidth]{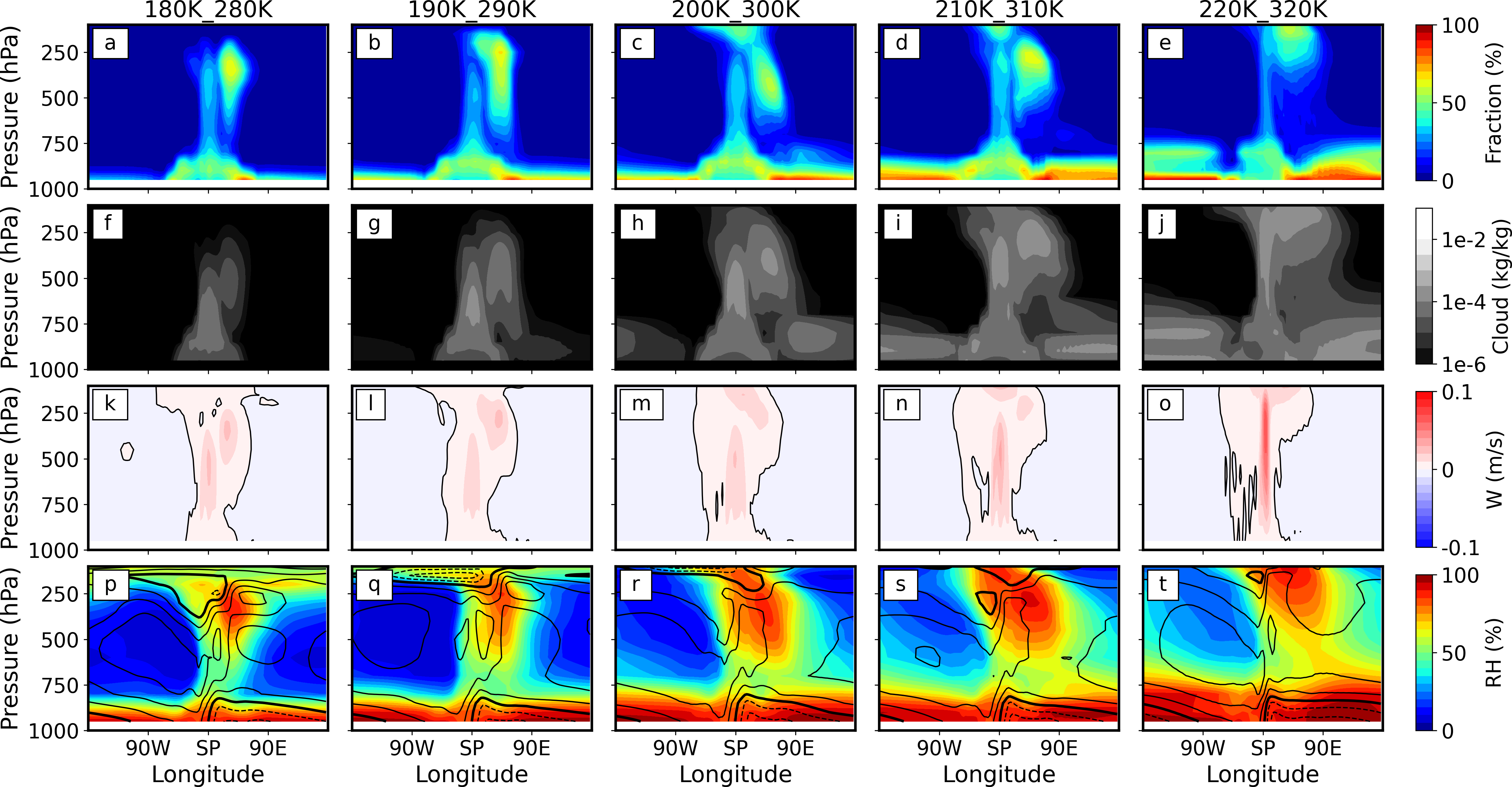}
    \caption{\textcolor{black}{Same as Figure \ref{fig:GCM_uni_no_rotation} for the group of GCM experiments of uniformly increasing surface temperature (Figure \ref{fig:SST_GCM}(a)) but with the effect of rotation. The rotation period ($=$~orbital period) is 30 Earth days.}}
    \label{fig:GCM_uni}
\end{figure}

\begin{figure}[t]
    \centering
    \includegraphics[width=0.9\textwidth]{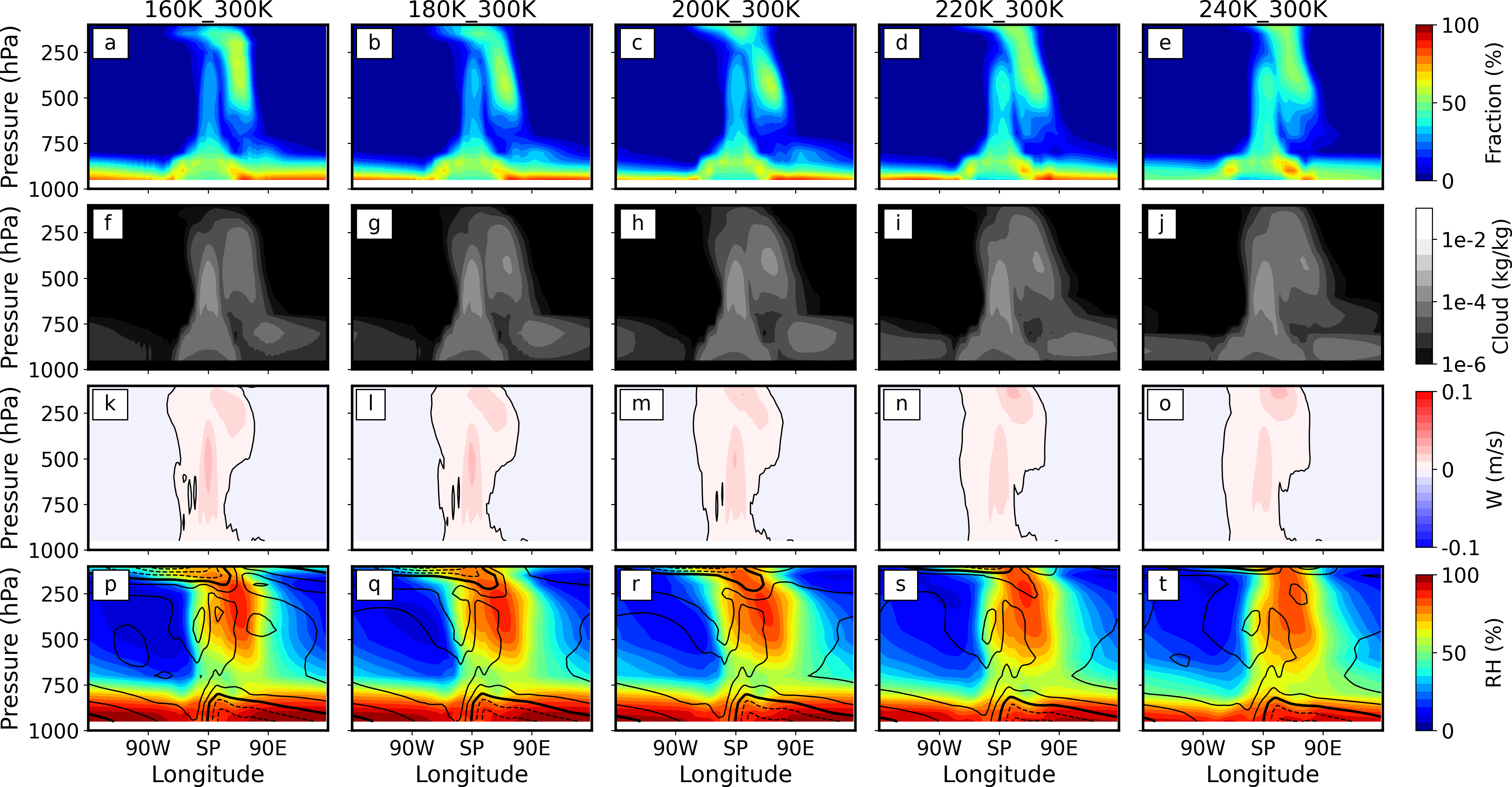}
    \caption{\textcolor{black}{Same as Figure \ref{fig:GCM_fixm_no_rotation} for the GCM experiments of changing the day-night surface temperature contrast (Figure \ref{fig:SST_GCM}(b)) but with the effect of rotation. The rotation period ($=$~orbital period) is 30 Earth days.}}
    \label{fig:GCM_fixm}
\end{figure}

\begin{figure}[t]
    \centering
    \includegraphics[width=0.9\textwidth]{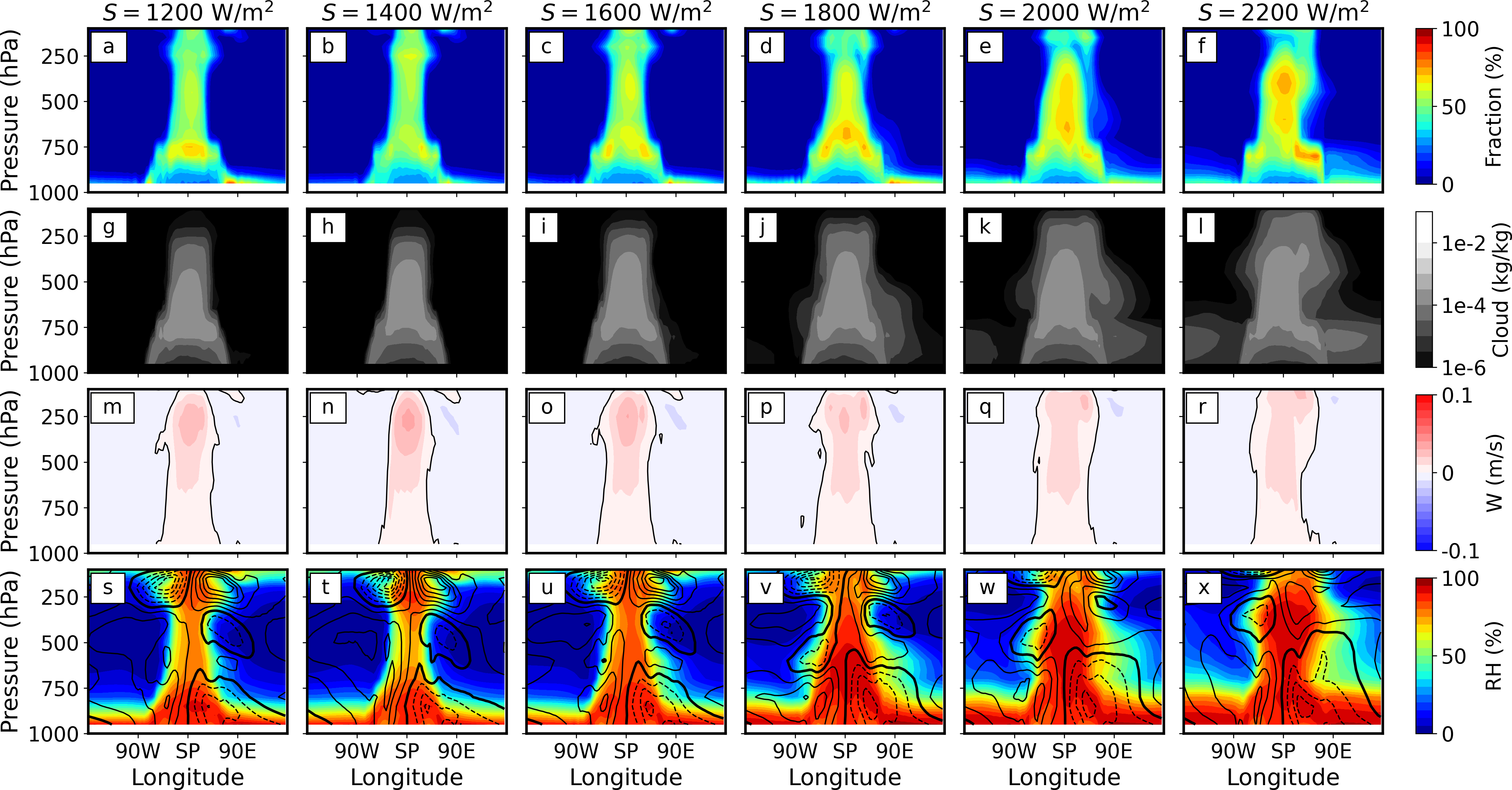}
    \caption{\textcolor{black}{Same as Figure \ref{fig:GCM_uni_no_rotation} but for the GCM experiments coupled to a 50-m slab ocean with varying stellar flux and with the effect of rotation (see Figure \ref{fig:SST_GCM}(c) for surface temperature). The rotation period ($=$~orbital period) is 30 Earth days.}}
    \label{fig:GCM_S}
\end{figure}

\begin{figure}[t]
    \centering
    \includegraphics[width=0.6\textwidth]{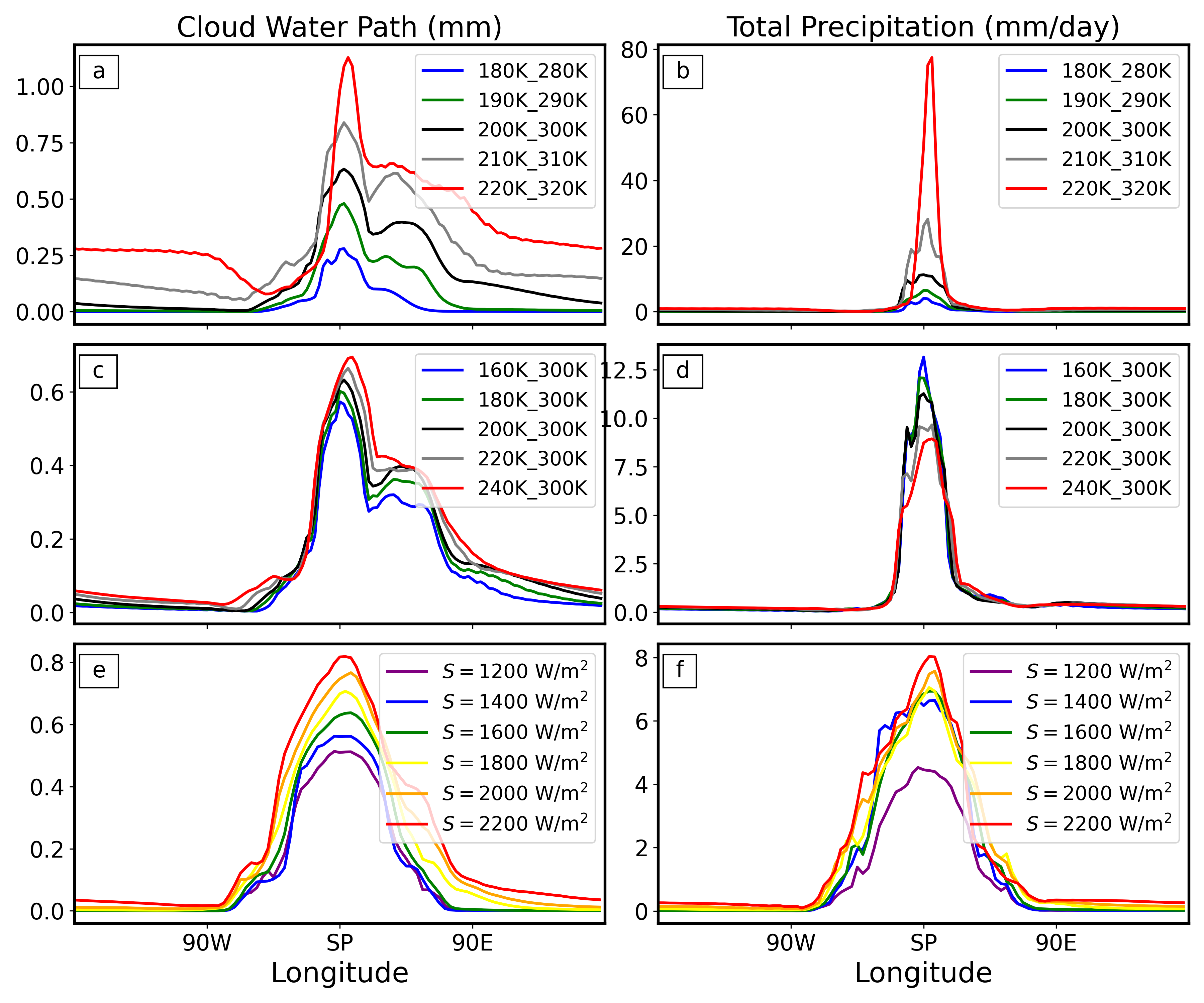}
    \caption{\textcolor{black}{Same as Figure \ref{fig:GCM_1D_no_rotation}, but for the GCM experiments with the effect of rotation. Panels (a-b), (c-d), and (e-f) are for the GCM experiments with uniformly increasing surface temperature (specified), reducing day-night surface temperature contrast (specified), and increasing stellar radiation coupled to a slab ocean, respectively. The rotation period ($=$~orbital period) is 30 Earth days.}}
    \label{fig:GCM_1D}
\end{figure}

\textcolor{black}{To better reflect the reality, we also did three groups of GCM experiments including the effect of planetary rotation. The first and second groups have the same configurations as the above GCM experiments of uniformly increasing surface temperature (Figure \ref{fig:SST_GCM}(a)) and reducing surface temperature contrast (Figure \ref{fig:SST_GCM}(b)), except for that rotation is now included. In the third group, the surface is coupled to a 50-m slab ocean, rather than with fixed surface temperatures. The results of these three groups of experiments are shown in Figures~\ref{fig:GCM_uni}, \ref{fig:GCM_fixm}, \ref{fig:GCM_S}, and \ref{fig:GCM_1D}. Again, no experiment of fixed radiative cooling was performed. }

In the first group of the GCM experiments \textcolor{black}{with rotation}, the cloud fraction, cloud water concentration, and vertical velocity generally become larger when the surface is uniformly warmed (Figure~\ref{fig:GCM_uni}(a-o)). But, the width of the ascending branch of the large-scale circulation decreases, as clearly seen in Figure~\ref{fig:GCM_uni}(k-o). These trends are roughly consistent with those found in the corresponding group of GCM experiments without rotation (Figure~\ref{fig:GCM_1D_no_rotation}(a \& b)). 

In the second group of the GCM experiments \textcolor{black}{with rotation}, when the day-night surface temperature contrast is reduced, \textcolor{black}{both cloud water path and cloud width increase}, the convective cloud area becomes wider (Figure~\ref{fig:GCM_fixm}(a-j)), but the strength of the ascending branch of the large-scale circulation becomes smaller (Figure~\ref{fig:GCM_fixm}(k-o)). Surprisingly, these trends are also qualitatively consistent with those found in the corresponding group of the GCM experiments (Figure~\ref{fig:GCM_1D_no_rotation}(c \& d)). 

In the third group of the GCM experiments \textcolor{black}{with rotation}, when the stellar flux is increased, \textcolor{black}{it is clear to see that the surface temperatures of both dayside and nightside increase, and the night-side warming is larger than the day-side warming (\ref{fig:SST_GCM}(c)).} The width of the ascending branch of the large-scale circulation increases (Figure~\ref{fig:GCM_S}(m-r)). The strength of the large-scale circulation shows a \textcolor{black}{slightly weaker} trend. Cloud fraction (Figure~\ref{fig:GCM_S}(a-f)), cloud water concentration (Figure~\ref{fig:GCM_S}(g-l)), and relative humidity (Figure~\ref{fig:GCM_S}(s-x)) increase significantly with increasing the stellar flux. 

When uniformly increasing the surface temperature (Figure~\ref{fig:GCM_1D}(a,b)), both the cloud water path and the precipitation show higher peaks, but the changes in width are not significant. The night-side clouds increase a lot, which is not seen in the CRM results (Figure~\ref{fig:state_1D}(a)). When reducing the day-night surface temperature contrast (Figure~\ref{fig:GCM_1D}(c,d)), the changes in both cloud water path and precipitation are small, which is consistent with the CRM results (Figure~\ref{fig:state_1D}(c,d)), but the peak of cloud water path increases with the reduction of temperature contrast, which is opposite to the CRM results. When increasing the stellar radiation, the surface temperature increases and the day-night surface temperature contrast decreases (Figure~\ref{fig:SST_GCM}(c)), and meanwhile both the cloud water path and precipitation increase in strength and become wider in horizontal extent, which is consistent with the previous GCM results such as in \cite{yang2013stabilizing}.

\section{Discussion} \label{sec:discussion}

\subsection{CRM Results versus GCM Results}

\begin{deluxetable*}{lcccc}[b]
\tablenum{2}
\tablecaption{Summary of the Responses of the Clouds and Ascending area of the Atmospheric Circulation in the Different Experiments}\label{tab:sum}
\tablewidth{0pt}
\tablehead{
\multirow{2}{*}{Models and Experiments} & \multicolumn{2}{c}{Cloud} & \multicolumn{2}{c}{Ascending area}
\\
&\colhead{Max CWP} & \colhead{Width} & \colhead{Width} & \colhead{Intensity} 
}
\startdata
\multicolumn{5}{l}{2D CRM without rotation}   \\
\qquad Uniform Surface Warming                         & $\Uparrow$   & \textcolor{blue}{$\Downarrow$} & $\Downarrow$  & $\Uparrow$       \\
\qquad Reducing day--night temperature contrast & \textcolor{blue}{$\Downarrow$}   & $\Uparrow$ & $\Uparrow$   & $\Downarrow$       \\
\qquad Enhancing radiative cooling rate                & $\Uparrow$   & $\Uparrow$ & $\Uparrow$   & $\Uparrow$       \\ \hline
\multicolumn{5}{l}{3D GCM without rotation} \\
\qquad Uniform Surface Warming                         & $\Uparrow$   & \textcolor{blue}{$\Uparrow$} & $\Downarrow$   & $\Uparrow$       \\
\qquad Reducing day--night temperature contrast & \textcolor{blue}{$\Uparrow$}   & $\Uparrow$ & $-$   & $\Downarrow$       \\ \hline
\multicolumn{5}{l}{3D GCM with rotation}        \\
\qquad Uniform Surface Warming                         & $\Uparrow$   & \textcolor{blue}{$\Uparrow$} & $\Downarrow$   & $\Uparrow$       \\
\qquad Reducing day--night temperature contrast & \textcolor{blue}{$\Uparrow$}   & $\uparrow$ & $\uparrow$   & $\Downarrow$       \\
\qquad Increasing solar fluxes                         & $\Uparrow$   & $\Uparrow$ & $\Uparrow$   & $\Downarrow$       \\ \hline
\enddata
\tablecomments{CWP is a short for cloud water path. $\Uparrow$ or $\Downarrow$ represents a significant increase or decrease, $\uparrow$ represents a slight increase, $-$ represents almost no change. Arrows with blue color represent that the signs of the trends are   opposite among the experiments.}
\end{deluxetable*}

\textcolor{black}{
The similarities and differences in the results between 2D CRM and 3D GCM are summarized in Table \ref{tab:sum}, within which the trends of the magnitude of cloud water path, the width of substellar clouds, the area of upwelling branch of the global Walker circulation, and the strength of the global Walker circulation are briefly listed for the experiments of 2D CRM without rotation, 3D GCM without rotation, and 3D GCM with rotation. Most of the metrics have similar trends when uniformly increasing the surface temperature or reducing the day--night surface temperature contrast. But, there are two exceptions. One is that the cloud band becomes narrower in the 2D CRM under uniformly increasing the surface temperature, but it becomes wider in the 3D GCM experiments of both with and without rotation. The other one is that the magnitude of the cloud water path decreases in the 2D CRM under reducing the day--night surface temperature contrast, but it increases in the 3D GCM experiments of both with and without rotation. In the 2D CRM, the trend of the convective cloud area is well consistent with the trend in the area of the upwelling branch of the Walker circulation, but in the GCM, the influence of the equatorial superrotation on the extension of the high-level clouds is so strong that the trends of these two fields are not always synchronous. Moreover, the magnitude of cloud water path in the 2D CRM is about one order higher than that in the 3D GCM, but the width of the dayside clouds in the 2D CRM is much narrower than that in the 3D GCM (Figure \ref{fig:state_1D} versus Figures \ref{fig:GCM_1D_no_rotation} \& \ref{fig:GCM_1D}). These differences can have significant influences on the strength of the cloud albedo feedback. Furthermore, there are more eddies and turbulent structures in the 3D GCM experiments due to the existence of waves, eddies, and instabilities, whereas all of these are not included in the 2D CRM simulations. Moreover, due to the simple boundary layer scheme used in the CRM simulations, the depth of the near-surface branch of the global Walker circulation is much larger than that in the GCM, $\approx$6--10 km versus $\approx$2--5 km. Based on previous studies, a shallow boundary layer may be more reliable \citep{Wordsworth_2015,koll2016temperature}.
}
\subsection{The Effect of Planetary Rotation}

\textcolor{black}{In our 2D CRM simulations, planetary rotation is excluded. In our 3D GCM simulations, both with and without rotation experiments were performed. So, these GCM experiments can be used to understand the effect of planetary rotation on the results (Figure \ref{fig:GCM_uni_no_rotation} versus Figure \ref{fig:GCM_uni}, Figure \ref{fig:GCM_fixm_no_rotation} versus Figure \ref{fig:GCM_fixm}, and Figure \ref{fig:GCM_1D_no_rotation} versus Figure \ref{fig:GCM_1D}), although only a slow rotator (30 Earth-days rotation period) is considered here. Several clear points can be found in the comparisons. (1) Including planetary rotation does not change the overall picture of the climate: deep convective clouds and robust precipitation over the substellar region and the global Walker-like circulation between dayside and nightside. (2) The magnitudes of cloud water path and rainfall do not change much. (3) The mean zonal winds in the free troposphere of the 3D GCM with rotation are in the direction from west to east, or called equatorial superrotation (see the contour lines in the bottom panels of Figures~\ref{fig:GCM_uni}, \ref{fig:GCM_fixm}, and \ref{fig:GCM_S}). The superrotation is maintained by eddy momentum transport from extra-tropics to tropics, ultimately driven by the uneven stellar flux distribution between the dayside and the nightside  \citep{showman2010matsuno,showman2011equatorial,pierrehumbert2019atmospheric}. The eddy momentum transport and equatorial superrotation are absent in the 3D GCM simulations without rotation as well as in the 2D CRM simulations. (4) The asymmetry between the west side and the east side of the substellar point is large in the 3D GCM experiments with rotation, such as that there are more clouds on the east side than on the west side, and the high relative humidity region also tilts towards the east side. This is mainly due to the combined effects of waves and the equatorial superrotation. But, in 3D GCM simulations without rotation as well as in the 2D CRM simulations, the symmetry between the west side and the east side is almost perfect.}

\textcolor{black}{Moreover, different rotation rates have a significant effect on the clouds and climatology of tidally locked planets. A faster rotation rate means a smaller Rossby deformation radius and a smaller Rhine length, which lead to the classification of “slow”, “Rhine”, or “fast” rotators, as described in \cite{Haqq-Misra2018demarcating}. For the first two classes of planets, the hemispheric large-scale Walker-like circulation is maintained and a strong convergence zone near the substellar area is expected; the convergence zone for “Rhine” rotators is more extended in the meridional direction due to planetary-scale turbulent flows. The day-to-night overturning circulation would eventually break for fast rotators, and the day-night temperature contrast would be weaker in the equatorial area while the equator-to-pole temperature difference would be larger. These different patterns in the atmospheric circulation are expected to result in different cloud distributions and different strengths of the cloud albedo feedback, which will be an interesting topic to be addressed in future subsequent CRM research. Our 2D CRM and 3D GCM simulations here mostly address the slow rotators, focusing on the response of the aggregated deep clouds (or the convergence zone) to different forcings.}

\subsection{The Spatial Pattern of Radiation Cooling Rate}

\begin{figure}[t]
    \centering
    \includegraphics[width=\textwidth]{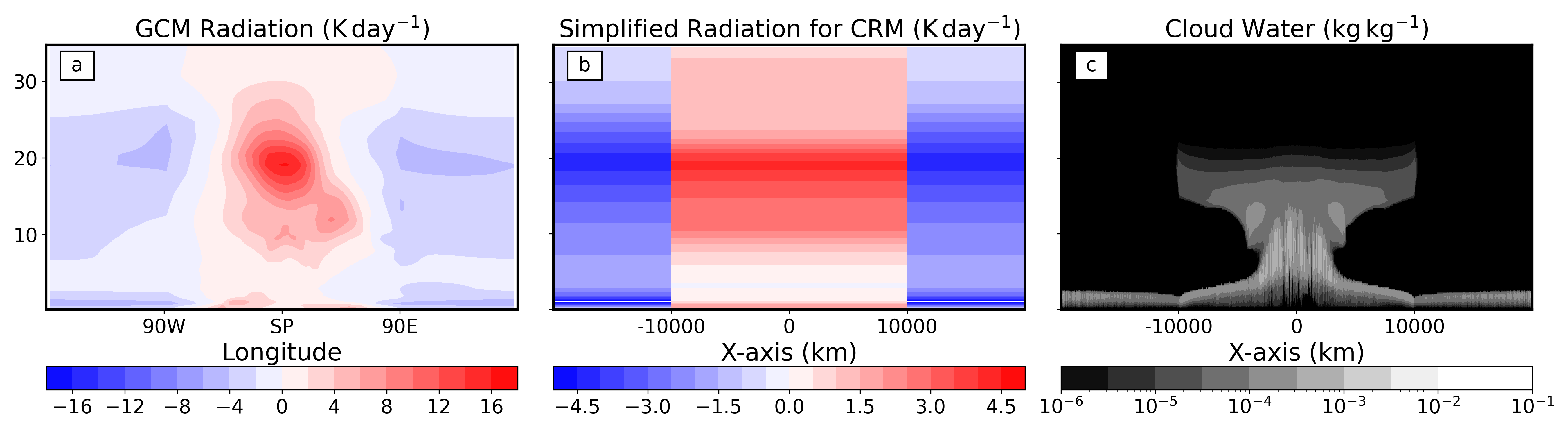}
    \caption{\textcolor{black}{The non-uniform radiation heating rates and the result for an CRM experiment. (a) Meridional-mean distribution of radiation heating rate in the 3D GCM experiment with surface temperatures ranging from 210 to 310 K. (b) 2D distribution of the radiation heating rate used in new CRM experiment. (c) Cloud water distribution in the CRM equilibrium state with the radiation heating rate of (b).}}
    \label{fig:vary_rad}
\end{figure}

\textcolor{black}{In our CRM experiments, we prescribed a spatially uniform radiation cooling rate in the simulated domain for simplicity. However, simulations using GCM have shown that the radiation distribution is non-uniform (Figure \ref{fig:vary_rad}(a)). Radiation cools the atmosphere on the nightside and heats the atmosphere on most parts of the dayside. The fine structure of the distribution is associated with the distributions of stellar radiation, water vapor concentration, air temperature, and cloud water content. As a test, we calculate the horizontal-mean radiation heating/cooling rates separately for the dayside and the nightside (Figure \ref{fig:vary_rad}(b)) and prescribe the results as the 2D distribution of radiation in the CRM experiment with surface temperatures ranging from 260 to 310 K. The resulting cloud water still shows the global Walker-like circulation with convective clouds covering the substellar area (Figure \ref{fig:vary_rad}(c)). Compared with the CRM experiment with uniform radiative cooling and the same surface temperature (Figure \ref{fig:group_1_state}(h)), the night-side low-level cloud fraction increases, and the high-level clouds on the dayside widen a lot. The increase of low-level cloud fraction on the nightside is due to the enhanced radiative cooling rate.}


\subsection{Cloud Albedo}


 \textcolor{black}{A commonly used metric for evaluating the radiation effect of clouds on climate is by calculating its albedo. However, since our model is not coupled with a realistic radiation transfer model, we cannot explicitly calculate the variables related to radiation processes such as cloud albedo. A calculation of cloud albedo will take many assumptions, including the phase (liquid phase, ice phase, or mixed) of cloud drops, the radius distribution of cloud drops, the vertical overlapping of different layers of clouds, etc. With so many unknown parameters, it’s hard to process an offline, reasonable calculation of the cloud albedo. But, we can expect that if any other conditions are the same, the cloud water path, which is a vertical integral of the cloud water content, would have a positive effect on the cloud albedo, meaning a higher cloud water path leads to a higher cloud albedo. Also, if the cloud width is the only varying factor, the cloud albedo will increase as the cloud width increases. In the future, we hope to couple the CRM with an online radiative transfer model to better address the cloud radiation effects.}


\subsection{Comparisons with Previous CRM Results}

\textcolor{black}{Before this work, three studies used cloud-resolving models to simulate tidally locked habitable planets \citep[e.g.,][]{zhangetal2017surface, sergeev2020atmospheric, Lef_vre_2021}. These studies employed different models, different dynamical cores, different physical packages, different planetary parameters (such as rotation period), and different initial/boundary conditions, which make it extremely hard or impossible to directly compare their results with the simulations here. Instead, we briefly summarize their studies separately and then address the common findings.}

\textcolor{black}{\cite{zhangetal2017surface} simulated the substellar clouds using the Weather Research Forecasting model (WRF) under a small domain of about 1000 km by 1000 km with a resolution of 3 km. The initial and boundary conditions were from the 3D GCM--CAM3. They found that there are strong spatial variabilities in clouds and surface temperature in the region near the substellar point. \cite{sergeev2020atmospheric} examined the effect of different convection parameterization schemes on the simulated climates of two exoplanets, Trappist-1e and Proxima Centauri b. They found that moist convection schemes play a key role in influencing the clouds, global atmospheric circulation, water cycle, and day-night temperature contrast. \cite{Lef_vre_2021} also employed the model of WRF but with modified physical packages (including radiative transfer and cloud/precipitation microphysics). Three different levels of stellar flux and two rotation periods were considered in their experiments; initial and boundary profiles of temperature, pressure, winds, and water were from the time averages of one 3D GCM; and the horizontal resolution was 1 km over a small domain of 250 km by 250 km. They found that the substellar region should be covered by convective clouds, but the stabilizing cloud albedo feedback as increasing stellar flux is weaker than that simulated in one 3D GCM. Here, we perform the simulation with a 2D global cloud-resolving configuration, covering an entire circle along the equator. This way of simulation means we do not need to interpolate or apply any boundary conditions from coarse-resolution models, thus avoiding the induced uncertainties both mathematically and scientifically. In our study, we separate the controlling factors of the overturning circulation and clouds into three factors (overall surface temperature, day-night surface temperature contrast, and radiation cooling rate). This separation helps us better evaluate the climate state of tidally locked planets. All the previous studies as well as our work here agree that convection is a key part of simulating the climate of tidally locked habitable planets, the substellar region should be covered by convective clouds, and the substellar clouds have strong temporal and spatial variations, although cloud properties (such as cloud fraction, cloud water path, cloud altitude, etc.) have significant differences between different models or different convection schemes.}

\section{Summary} \label{sec:sum}
The simplified 2D cloud-resolving model, SNAP, was employed to investigate the general characters of clouds and atmospheric circulation and the effects of varying surface temperature and infrared radiation on the convection and clouds of tidally locked habitable planets. The main results are summarized as follows:

\begin{enumerate}
    \item Stimulated by the centrally-peaked surface temperature pattern and the uniform radiative cooling, we obtain a global-scale Walker-like circulation and optically thick clouds in the substellar area on the dayside of the planet.
    
    \item When uniformly increasing the surface temperature, the ascending area narrows mainly due to the increased MSE contrast between the substellar area and the nightside. The width of high-level anvil clouds also decreases, mainly due to the increase of atmospheric stability. But, the strengths of convection and large-scale circulation increase, due to the increased MSE difference between the near surface and the free troposphere.
    
    \item When the day-night surface temperature contrast is decreased, the convection and large-scale circulation become weaker in strength but wider in area. These could be explained by the decrease of the day--night MSE contrast. 
    
    \item When the radiative cooling rate is increased, convection and large-scale circulation become stronger, and the convection area and the width of high-level anvil clouds increase. These can be explained based on global energy budget and radiation-driven downwelling and divergence.
    
    \item The results of 2D cloud-resolving experiments are broadly similar to those in corresponding 3D GCM experiments, \textcolor{black}{such as the existence of deep convective clouds over the substellar region, although there are several significant differences, such as the magnitude of cloud water path, cloud width, and their trends (see Table \ref{tab:sum}).} This bolsters the use of GCMs in exoplanetary climate studies, \textcolor{black}{at least in the qualitative trend of clouds under increasing stellar flux.}
    
    \item \textcolor{black}{The magnitude of cloud water path in CRM is about one order larger than that in GCM, but the width of the substellar cloud region in CRM is much narrower than that in GCM. These differences could significantly influence planetary albedo and its spatial pattern.}

\end{enumerate}

The weaknesses of the cloud-resolving simulations shown here are: the simulation domain is 2D along the equator rather than a real 3D sphere, and the effect of the Coriolis force is not included, the radiative transfer module is so simple that realistic radiative transfer is absent, the cloud microphysical processes are very simple, and the surface temperature is fixed rather than coupled to the atmosphere or the ocean. Future 3D simulations including realistic radiative transfer and a coupled slab ocean or dynamical ocean will be better to represent the climate system of a tidally locked habitable planet.

The results of the three groups of CRM experiments suggest that cloud water amount and cloud coverage would likely increase when the stellar flux is increased if realistic radiative transfer were included in the model. This can cause a stabilizing cloud feedback and make the inner edge of the habitable zone closer to the host star, as suggested by previous GCM simulations \citep[e.g.,][]{yang2013stabilizing,kopparapu2016inner}. Further work is required to confirm this using a 3D CRM with realistic radiative transfer and more complex cloud microphysical schemes.



\acknowledgments
We are grateful to Bolei Yang and Zhihong Tan for their helpful suggestions. \textcolor{black}{We thank Dorian S. Abbot for his very insightful and helpful suggestions towards the draft.} We also thank Xinyi Song for helping with the language revision. J.Y. acknowledges support from the National Natural Science Foundation of China (NSFC) under grants 42161144011 and 42075046.


\appendix


\begin{figure}[t]
    \centering
    \includegraphics[width=0.9\textwidth]{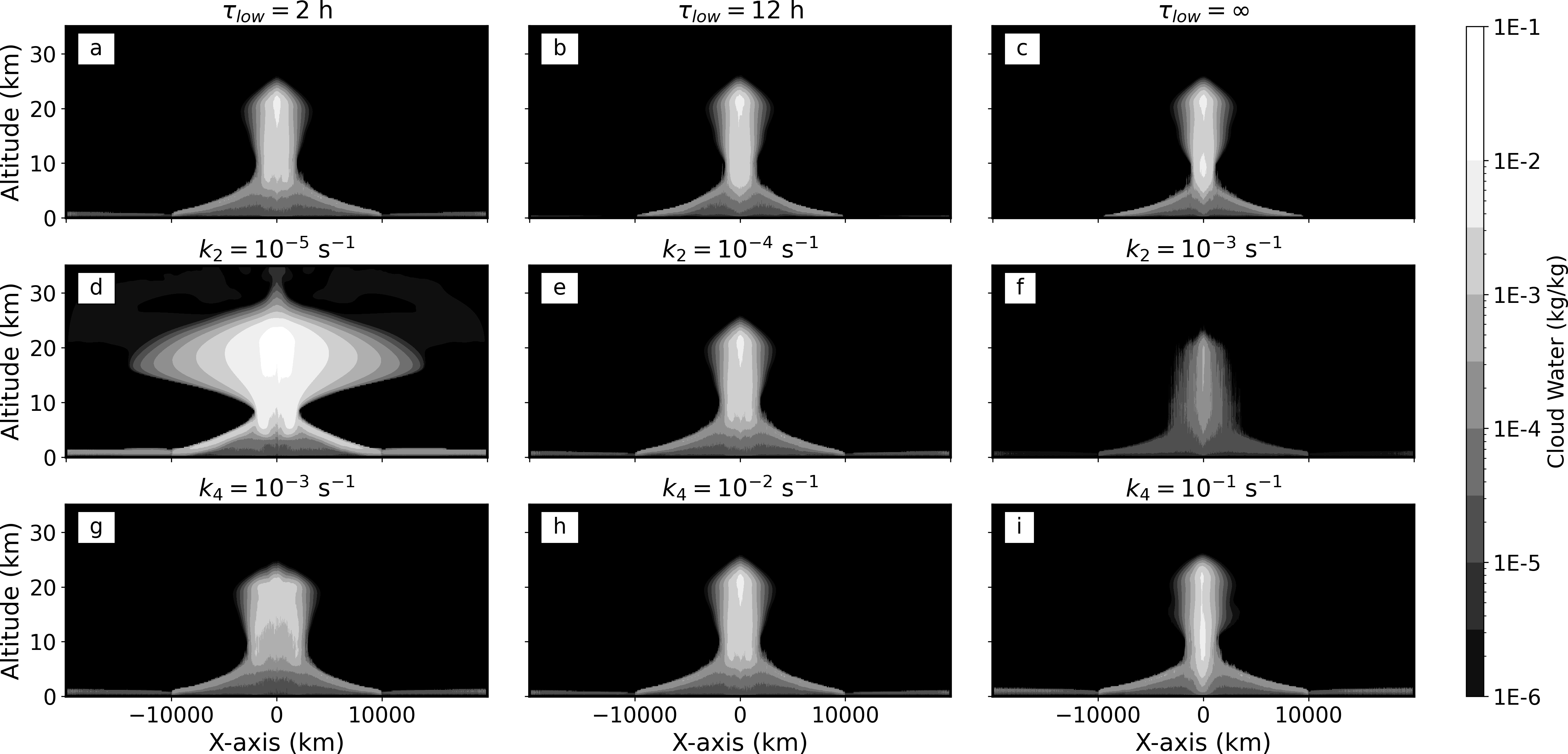}
    \caption{Cloud water concentration in the sensitivity tests of (a-c) varying the surface friction time scale ($\tau_{low}$ in Equation~(\ref{eq:momentum_diss})), (d-f) varying the conversion rate from cloud water to precipitating water ($k_2$), and (g-i) varying the re-evaporation rate of rain droplets back to water vapor ($k_4$). Each panel represents an experiment, and its corresponding parameter is marked at the top of the panel.}
    \label{fig:ST}
\end{figure}


Here we perform three extra groups of sensitivity tests in order to test the effects of boundary layer friction and two \textcolor{black}{bulk physics} parameters. These results below suggest that boundary layer scheme and \textcolor{black}{bulk physics} parameters can strongly influence the simulated results of clouds, although these parameters do not completely change the basic feature of the large-scale circulation and convection. 

The effect of the boundary layer friction is tested in three experiments in which we set the time scale of friction-induced momentum damping as 2 hrs, 12 hrs, and infinite long (i.e., free-slip surface), respectively. A longer time scale means a weaker friction effect. The results of cloud water concentration are shown in Figure \ref{fig:ST}(a-c). As the friction weakens, the convective area and the ascending branch of the large-scale circulation become narrower, similar to that found in \cite{Liu2008explicitly}. Weaker surface friction leads to stronger near-surface horizontal winds and more centralized convergences, so the ascending branch becomes narrower. In the default experiment, the inflow winds have maximum strength at the level of about $1~\mathrm{km}$. In the free-slip experiment, the maximum winds reach the lowest level of the model, and the temperature inversion penetrates to the surface, leaving no space for night-side cloud formation; as shown in Figure~\ref{fig:ST}(c), there are no night-side clouds in this experiment.

The \textcolor{black}{bulk physics} processes in the model are described in Equations~(\ref{eq:micro_1})--(\ref{eq:micro_3}), in which $k_2$ means the cloud-to-precipitation conversion rate. A larger $k_2$ means a quicker conversion from cloud water to precipitating water. We change $k_2$ within two orders of magnitude, and the resulting cloud water concentration is shown in Figure \ref{fig:ST}(d-f). As $k_2$ increases, cloud water concentration decreases strongly, and the width of the high-level anvil clouds becomes narrower significantly, due to the direct effect of a faster transition from clouds to rainfall. However, the domain-average rainfall remains nearly the same in these experiments (figure not shown) because the latent heat release must balance the radiative cooling, which is fixed.

The re-evaporation rate of precipitating water ($k_4$) also affects the results. A larger $k_4$ means a quicker re-evaporation from rain droplets to water vapor under an unsaturated environment. We change $k_4$ within two orders of magnitude, and the resulting cloud water concentration is shown in Figure~\ref{fig:ST}(g-i). As $k_4$ increases, the convective area and the ascending branch of the large-scale circulation become narrower. The reason is that enhanced re-evaporation and related heat absorption reduce the magnitude of diabatic heating, suppressing convective activities, narrowing the ascending area, and reducing the convective cloud area.


\bibliography{sample63}{}
\bibliographystyle{aasjournal}


\end{document}